\documentclass[runningheads]{llncs}
\usepackage[english]{babel} 
\usepackage{ucs}
\usepackage[utf8x]{inputenc}
\usepackage[T1]{fontenc}
\usepackage{pslatex}
\usepackage{graphicx}
\usepackage{xspace}
\usepackage{float}
\usepackage[usenames]{color}
\usepackage{bsymb,b2latex}
\usepackage{boxedminipage}
\usepackage{ulem} % pour barrer du texte, ...
\usepackage[pdftex]{thumbpdf} 
\usepackage[bookmarks=true,bookmarksnumbered=true,colorlinks=false,urlcolor=blue,citecolor=blue,linkcolor=blue]{hyperref}

\definecolor{darkred}{rgb}{0.8,0,0}

\usepackage{./Kmelia}
 
\usepackage{./B2}

\protect

\protect
\spnewtheorem*{nota}{Note}{\bfseries}{\itshape}
\newif\ifwithcomments
\withcommentstrue % WARNING: line to comment to compile a non-commented version

\begin{document}

\title{Modelling and Analysing the Landing Gear System:\\
 a Solution with Event-B/Rodin}

% \titlerunning{}

\author{Pascal Andr\'e, Christian Attiogb\'e and
  Arnaud Lanoix}
%\authorrunning{}

% Pascal Andr\'e 
% \and Christian Attiogb\'e 
% \and Arnaud Lanoix} % \inst{1}

\institute{AeLoS Team\\
  LINA CNRS UMR 6241 - University of Nantes\\
%  2, rue de la Houssini\`ere\\
%  F-44322 Nantes Cedex, France\\
  \email{\{firstname.lastname\}@univ-nantes.fr}}

\maketitle

\begin{abstract}
  % -*- coding: utf-8 -*-
%
%
This paper presents a solution to the landing gear system case study
using Event-B and Rodin.  We study the whole system (both the digital
part and the controlled part). We use {\it feature augmentation} to
build an abstract model of the whole system and {\it structural
refinement} to detail more specifically the digital part. The required
safety properties are formalised and proved. We propose a specific
approach to deal with a family of reachability properties.  The
experimentations conducted during the study are supported by the Rodin
tools.  We show that the presented solution is systematic and it can
be applied to similar case studies.

\end{abstract}

%\textbf{memo : deadline = January 14, 2014 / 3 février}

\section{Introduction}
\label{sec:intro}
% -*- coding: utf-8 -*-
%
%-----------------------------
% Landing System ABZ'2014
% PA, CA, AL

This work reports on an answer to the landing gear system case study
submitted to the state/proof-based formal methods community, in the
scope of the ABZ conference.  Our motivation is to contribute to the
challenges of this real-life case study with Event-B, to share and to
compare experiences and knowhow with other competitors. Even if the
Event-B method have been widely used in many
academical~\cite{eventb-book-abrial2010} and industrial case
studies~\cite{mondex_butler_2008,lanoix:hal-00260577,DBLP:conf/sbmf/DamchoomB09,DBLP:journals/sttt/CansellMP09,eventb-book-abrial2010,spaceCraft_FathabadiButler-2011,DBLP:journals/tecs/MeryS13},
there are still some challenging questions about methods, liveness
properties management, reusable specification patterns, etc.  However
Event-B is recognised as a mature, well-researched formal development
method which is equipped with mature engineering tools such as
Atelier-B and Rodin.  Accordingly we decide to present a complete
solution using Event-B. By the way we have not only to answer to the
case study but also to explore solutions to the challenging questions
in Event-B.

%contribution
We address in our proposal the two main challenges that characterise
the considered case study viewed as representative of critical
embedded systems. The challenges are firstly the modelling of the
control part of the landing system and secondly the proof of the
safety requirements.  Accordingly the contribution of our work is
manyfold: \textit{i)} a complete abstract modelling of the control
digital part in interaction with its physical environment using a
stepwise refinement with feature augmentation; \textit{ii)} the proof
of some safety properties of the landing system; \textit{iii)} the
refinement of the digital part to take into account its composition
with two redundant modules as described in the requirements. Our
modelling approach is summarised in Fig.~\ref{fig:principle}; the
presentation of the article also follows the structure depicted in
this figure.

We consider that the reader is familiar with Event-B and we do not
introduce the method and its features in this document; but the
interested reader can consult~\cite{eventb-book-abrial2010} for a
detailed introduction to Event-B.

%structure of the article
The article is structured as follows.
Section~\ref{section:reqAnalysis} provides the main idea and the
working hypotheses of our solution to the landing gear system.  In
Sect.~\ref{section:abstratModel} we detail how the global abstract
model of the system have been built with feature augmentation and we
explain how we capture the requirements of the landing system.
Section~\ref{section:refinedModel} is devoted to the structural
refinement where we provide some details for the decomposition of the
global model and to the detailed specification of the digital part.
In Sect.~\ref{section:discussion} we discuss our solution; we give a
synthesis of the approach and how it can be reused in similar control
systems. We give some feedbacks on the experimentations achieved with
Rodin.  Finally, section \ref{section:conclu} concludes the article
and provides some perspective works.

\section{Analysing and Capturing the Requirements}
% Methode globale d'analyse du "problème"
\label{section:reqAnalysis}
% -*- coding: utf-8 -*-
%
%-------------
% analysis.tex
%-------------
% Methode globale d'analyse du "problème" et de sa modelisation
%

In the category of critical embedded system, the landing system is
essentially a control system with time constraints.
In such control systems the whole system is made of a controller and a
controlled physical environment which interact via sensors and
actuators. In the requirement document of the landing system the
controller is called the \textit{digital part}; the controlled
mechanical and hydraulic environment is called the \textit{physical
  part}. We use this terminology along the article. The digital part
monitors and controls the physical part; the sensors provide to the
digital part the information on the state of the physical part; the
actuators engage the actions of the controller on the physical part.

\begin{figure}[htbp]
	\centering
	\includegraphics[width=.7\linewidth]{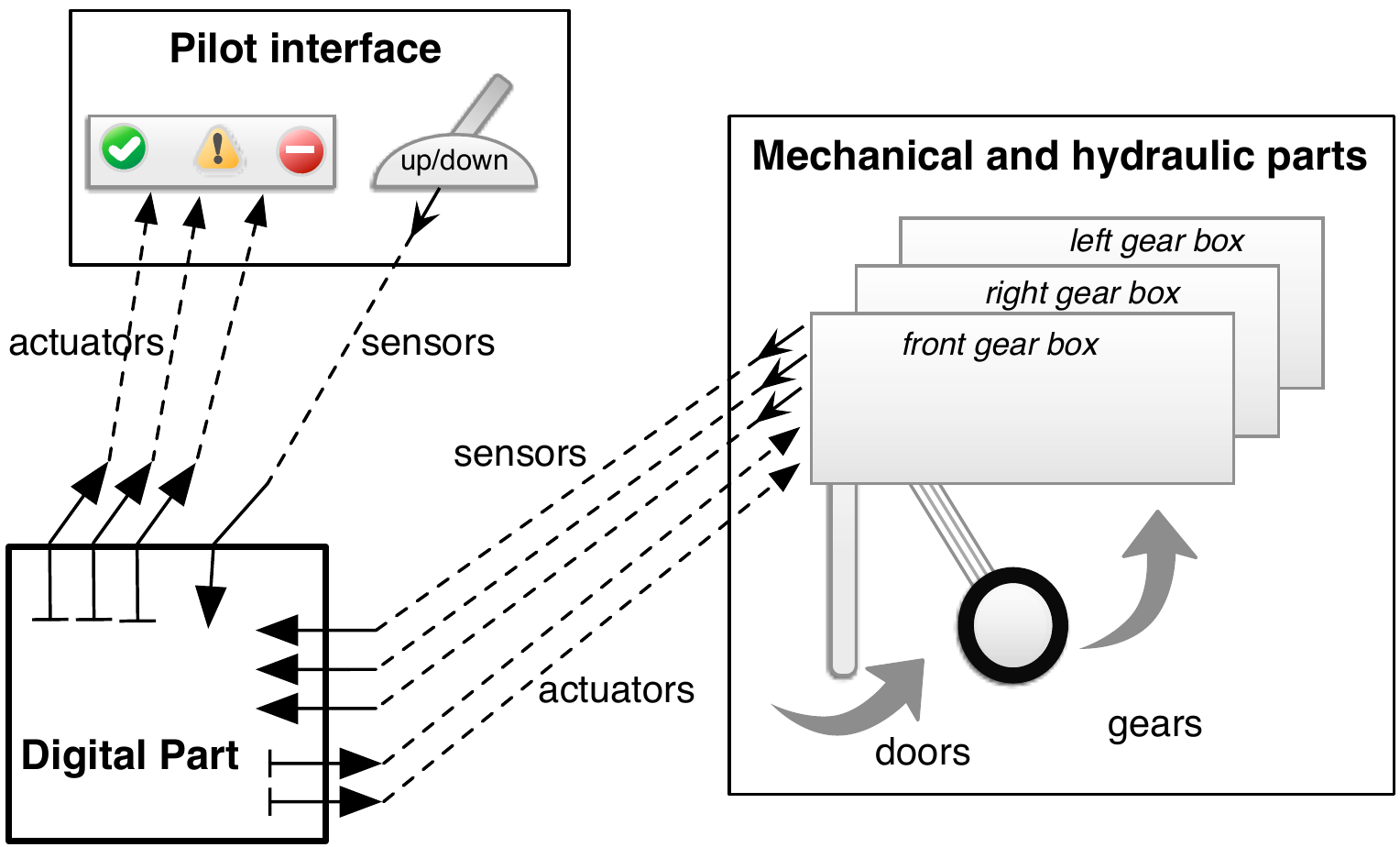}%width
	\caption{Global architecture of the Landing system}
	\label{fig:main:archi}
\end{figure}

\subsection{Analysing the landing system requirement}
The landing system requirements have already been structured in a way
that there is a clear separation between the digital part and the
physical part.  The physical part is mainly made of gears and doors;
but there is also a cockpit viewed as a control and supervisory
equipment; it is made of a handle driven by a human pilot. The digital
part is located between the handle and the physical part (see
Fig.~\ref{fig:main:archi}): the orders from the digital part to the
controlled (physical) part are originated from actioning the handle;
two main interactions are described; an interaction between the handle
and the digital part and an interaction between the digital part and
the controlled part.

\begin{figure}[htbp]
\centering
\includegraphics[width=0.7\linewidth]{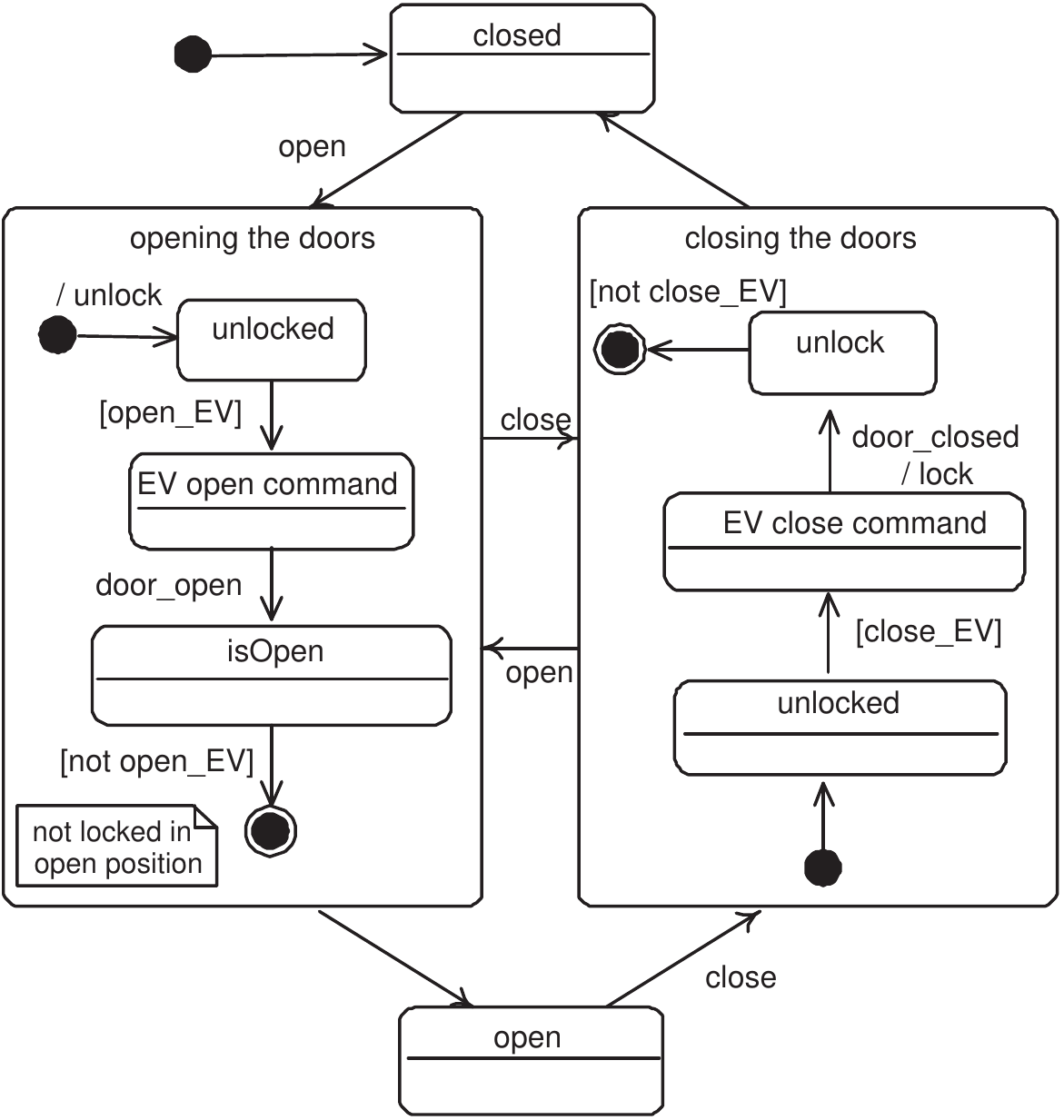}
\caption{The door state automaton}
\label{fig:door_beh}
\end{figure}

To understand and to capture precisely the details of the components
of the landing system and their behaviours, we have read several times
(and continuously along the modelling work) the requirement documents
and compare our views ; we have tested various approaches in the team:
\begin{itemize}
\item we have sketched many informal figures; 
\item we have drawn many state machines to capture the behaviour of
  doors, gears and the sequences of operations to extend or to retract
  the gears;
\item we have built UML diagrams to capture the data and the dynamic
  behaviour of both the control part and the physical devices; an
  example of a state diagram describing the doors behaviour is
  depicted in Fig.~\ref{fig:door_beh};
\item we have constructed preliminaries classical and Event-B
  specifications to identify data and behaviours;
\item we have used Z schemas to capture data structuring at this first
  level of analysis.
\end{itemize}

From the methodological point of view, it is greatly beneficial to
proceed in this way; indeed some facets of the requirements are easily
revealed by one approach or the other. This results in a accurate
understanding of the requirement and help in building a complete
abstract model.  A collection of drawing informal figures, UML
diagrams, Z schemas and additional works related to this step of our
work can be found in a dedicated
url\footnote{\url{http://www.lina.sciences.univ-nantes.fr/aelos/softwares/LGS-ABZ2014/}}.

We have classified the requirements listed in the case study document
(see page 18-19 of the requirement document) in several categories of
properties to be proved for the system.
\begin{description}
\item \textbf{Safety:} The requirements $R_2$, $R_3$, $R_4$ and $R_5$
  should be considered through safety properties.
\item \textbf{Liveness:} The requirements $R_1$ are related to
  liveness (reachability) properties.
\item \textbf{Nonfunctional:} The requirements $R_6$, $R_7$ and $R_8$
  (Failure mode requirements) are related to nonfunctional properties:
  management of time constraints.
\end{description}

The physical part is made of autonomous physical devices and its
global behaviour is a composition of the behaviours of the considered
devices.  We consider as in the requirement document that the physical
devices exist and we will not build them; the challenge deals with the
control part only (see page 2 of the requirement document). However we
have to consider an abstraction of the physical part in order to build
a global model of the landing system detailed enough to capture the
required properties.

The digital part is perceived first at the abstract interface level
and detailed (refined) progressively. A top-down approach as used in
Event-B is appropriate.

\begin{figure}[htbp]
	\centering
	\includegraphics[width=\linewidth]{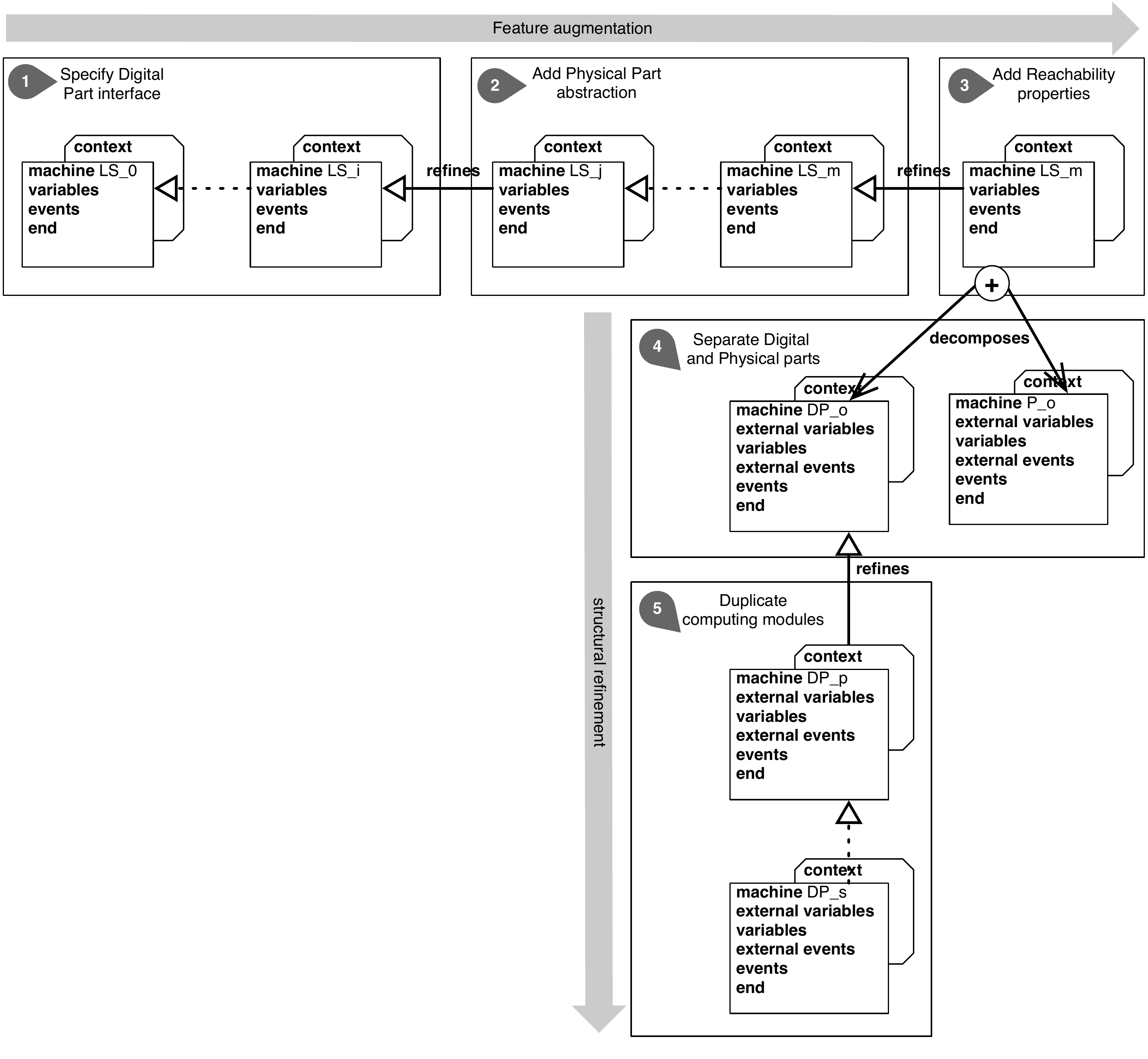}%width
	\caption{Principle of the modelling approach}
	\label{fig:principle}
\end{figure}

\subsection{Modelling Methodology}

The landing system is studied globally by considering both the digital
part and the physical part but only the digital part will be studied
in details. 

It has been established and demonstrated in several case
studies~\cite{DBLP:conf/sbmf/DamchoomB09,DBLP:conf/icfem/DamchoomBA08,eventb-book-abrial2010}
that complex systems can be constructed by combining horizontal
refinement (with feature augmentation in an abstract model) and
structural refinement (making the abstract structures more and more
concrete).

According to the complexity of the landing case study we use feature
augmentation to build the abstract model; indeed the landing system at
a first approximation contains the digital part and the physical part;
but each of these parts is also made of several structuring details
which can be introduced step by step.

From the point of view of methodology one question is how to determine
the starting point of the abstract model. There are different ways to
proceed with.
\begin{enumerate}
\renewcommand{\theenumi}{\textit{\roman{enumi}})}%
\item One can start by specify the physical part with a
  correct behaviour and progressively adding the digital part which
  preserves the behaviour of the physical part.
\item One can start with a desired digital only described
  by its interface with the (future) physical part, and progressively
  adding details on the physical part.
\item A conjoint construction of the digital part and the physical
  part as soon as the details of each part are needed to express
  either the properties or the needed structuring details of the whole
  system.
\end{enumerate} 

We have experimented with the three approaches.  It appears that the
second one is more relevant: we firstly specify the digital part
focusing on its interface with the physical part. Then, we have to
take into account the elements of the physical part before describing
some properties of the digital one (for example the reachability
properties). The interaction cycle \textsf{sense/decision/order} of
control systems requires the availability of a state space including
the data of both sides.

Consequently, following the system engineering approach of Event-B, we
will build an abstract model including first an abstract view of the
digital and then introducing the physical part; this is achieved using
a series of horizontal refinements (by feature augmentation).

We distinguish three main levels\footnote{note that each level is
  a set of very small steps of refinement} of refinement:
\begin{itemize}
        \item \textbf{Level 1 (external view of the digital part):}
          we focus on the interface of the digital part and
          the related events.  Inside this level, the features will be
          introduced progressively (interface, actions of the outgoing
          sequence, triplicated sensors, etc.)
        \item \textbf{Level 2 (introducing physical devices with their
            sensors):} the digital part is progressively linked to
          abstractions of the physical devices which simulate their
          real behaviours.
        \item \textbf{Level 3 (introducing reachability constraints):}
          properties related to time and reachability are dealt with; 
specific constructions are introduced for this purpose.
\end{itemize}

The result of these refinements is an overall abstract model of the
landing system, where many requirements are expressed and proved.
At \textbf{level 4}, this overall abstract model will be decomposed into
a model of the physical part and a model of the digital part; we use
the \textit{Event-B decomposition} approach to manage this step of the
process.

The \textbf{level 5 (duplicating the computing modules)} details the
digital part with a series of vertical refinements (called
\textit{structural refinements}). We distinguish here two main levels
of refinement.
\begin{itemize}
\item \textit{Introduction of the two computing modules:} the
  refinement at this level enables us to model the internal structure
  of the digital part, the two redundant modules are introduced. This
  step is structured with small refinement steps where we consider one
  category of the interface variables at a time: the inputs and then
  the outputs variables.
\item \textit{Modelling of the internal behaviour of the computing
    module:} in the current stage of our work, the last step of the
  structural refinement is the modelling of the internal behaviour of
  each computing module.
\end{itemize}

In the remaining sections of the article we detail each level listed
above. 

%===============================================

\section{Building an Abstract Model of the Landing System with Feature Augmentation}
% Construction du modele abstrait
\label{section:abstratModel}
% -*- coding: utf-8 -*-
%
%-----------------------------
% Landing System ABZ'2014
% 

The method we used is to start with a rough abstract model with few
desired properties and to incorporate little by little more details
until to reach an almost complete model of the whole system comprising
both the digital part and the controlled physical part.

\subsection{Stepwise construction of the abstract model}
\label{section:stepwiseAbsM}
\begin{figure}[htbp]
\centering
\includegraphics[width=.7\linewidth]{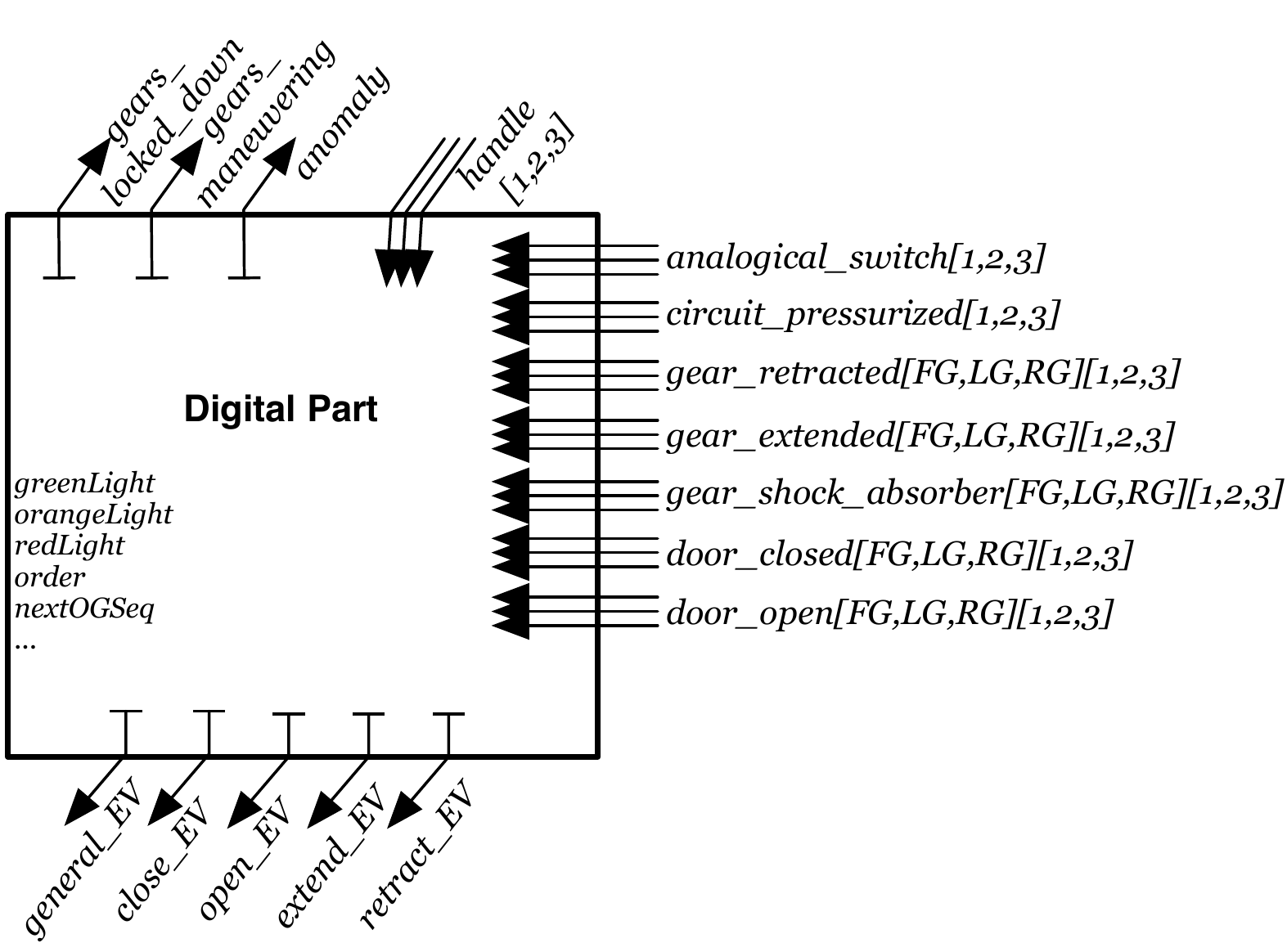}
\caption{The interface of the digital part}
\label{fig:interfaceAbstractModel}
\end{figure}

The requirement document is helpful (see page 7 of the document) to
identify the different variables at the interface between the digital
part and the physical part (Fig.~\ref{fig:interfaceAbstractModel});
they are the input and output variables introduced in pages 6 and 7 of
the requirement document.

Therefore, we distinguish three categories of variables at the
interface of the digital part.
\begin{itemize}
\item the \textit{input} variables: $handle$, $gear\_extended$,
  $gear\_retracted$, $analogical\_switch$,
 $door\-\_closed$, $door\_open$, etc.,
\item the  \textit{order output} variables: $general\_EV$,
  $close\_EV$, $open\_EV$, $extend\_EV$ and
  $retract \_EV$ which are used to send orders to the physical
  part and
\item the  \textit{state output} variables: $gear\_locked\_down$,
  $gears\_manoeuvring$ and $anomaly$ which are used to
  interact with the pilot interface.
\end{itemize}

We have identified the following basic enumerated sets: 
\begin{itemize}
\item $HSTATE = \{hDown, hUp\}$, which denotes the state of the
  $handle$ variable;
\item $AnalSWSTATE = \{openSW, closedSW\}$ which is used to model the
 $analogical \_switch$ variable;
\item $DOOR = \{FD,
RD, LD\}$ which denotes the front, the right and the left doors; 
\item $GEAR = \{FG, LG, RG\}$ which denotes the front, the right and the left
gears.
\end{itemize}

The input variables are triplicated, as indicated in the requirement
document. They are modelled with a type $TRIPLE =\{1,2,3\}$ used as an
index of the function variables; their descriptions are as follows:
\begin{center}
\begin{boxedminipage}{10cm}
$handle ∈ TRIPLE → HSTATE$\\
$analogical\_switch ∈ TRIPLE → AnalSWSTATE$\\
$gear\_extended ∈ (TRIPLE×GEAR) → BOOL$\\
$gear\_retracted ∈ (TRIPLE×GEAR) → BOOL$\\
$door\_closed ∈ (TRIPLE × DOOR) → BOOL$\\
$door\_open ∈ (TRIPLE × DOOR) → BOOL$\\
$\cdots$
\end{boxedminipage}
\end{center}
For instance with the function variable $handle ∈ TRIPLE → HSTATE$ we
capture precisely that $handle_i \in \{hDown, hUp\}$ with $i \in
\{1,2,3\}$.

\medskip 
The state output variables are modelled as follows:
\begin{center}
\begin{boxedminipage}{10cm}
$gears\_locked\_down  ∈ BOOL$\\
$gears\_maneuvering  ∈ BOOL$\\
$anomaly  ∈ BOOL$\\
$\cdots$
\end{boxedminipage}
\end{center}

\medskip 
The order output variables are modelled as follows:
\begin{center}
\begin{boxedminipage}{10cm}
$general\_EV  ∈ BOOL$\\
$close\_EV  ∈ BOOL$\\
$open\_EV  ∈ BOOL$\\
$\cdots$
\end{boxedminipage}
\end{center}

\medskip Additionally to these three categories of the interface
variables, we have some \textit{internal} variables; they are used
inside the controller. For example, to manage the state of the lights
which indicate the position of the gears and doors to the pilot, we
have $greenLight$, $orangeLight$, $redLight$.  They are bound to the
output state variable $gears\_locked\_down$ with an invariant
predicate.  In the following section we have other example of such
variables.

\subsubsection{Handling normal mode and anomaly}
Two modes are distinguished: a \textit{normal} mode where the system
is controlled digitally and its behaviour is correct if there is no
anomaly detected in the system otherwise a permanent failure is
observed; an \textit{emergency} mode where the system is controlled
analogically.  Only the normal mode is in the scope of the case study
(see page 1 of the requirement document).

Accordingly the boolean state output variable \textit{anomaly} is used
to denote that an anomaly has been detected or not (either
$anomaly=TRUE$ or $anomaly=FALSE$).  This variable is used to raise an
anomaly and also to distinguish the two modes in the specification of
properties and in the event guards.

One challenge of this case study is the construction of
an abstract model which is as faithful as possible with the
requirements. One of the properties of the landing system is
\textit{at the same time doors cannot be closed and open}. This is
captured by the following predicate into the invariant of the building
model:
\begin{center}
\begin{boxedminipage}{10cm}
$anomaly = FALSE ⇒ ran(door\_closed) ∩ ran(door\_open) = ∅ $
\end{boxedminipage}
\end{center}

\subsubsection{Deriving events from the action sequences}
The digital part of the system controls the hydraulic devices
according to the pilot orders and also to the mechanical devices
positions (page 13 of the requirement document). This is achieved
according to two specific action sequences: a landing action sequence
and a retraction action sequence. An internal variable $order ∈ HSTATE$
is used to pass on the order issued by actioning the handle.  The
variable $order$ is an example of \textit{internal} variable.

Moreover we have to take into account the requirements $R_2$ and $R_3$
(page 18 of the requirement document) which give the conditions of the
orders (\textsf{UP} or \textsf{DOWN}) that can initiate the landing
and retraction sequences and the conditions to enable them.

A thorough analysis of the two action sequences (outgoing sequence and
retraction sequence) of the landing system helps us to capture the
behaviour of the digital part. Even if they are nested each sequence
is analysed precisely; it is made of a sequence of transition from
state to state; each sequence is started as the effect of an action on
the handle by the pilot. The remaining transitions in the sequence are
mainly the orders from the digital part to the physical part, provided
that some conditions described in the requirement document are
established; therefore we are able to build a set of events (with
conditions-actions) that describe as Event-B guarded events, the
outgoing sequence and the retraction sequence.
 
In order to control perfectly the evolution of the outgoing sequence
we use a variable $nextOGseq$ which indicates in the event guards the
next step in the outgoing sequence. The variable is updated in the
body of the events.

\subsubsection{Modelling the outgoing sequence}
\label{page:outgoingSeq}
We now describe more precisely the outgoing sequence to illustrate our
approach.

\begin{figure}[htbp]
\centering
\includegraphics[width=\linewidth]{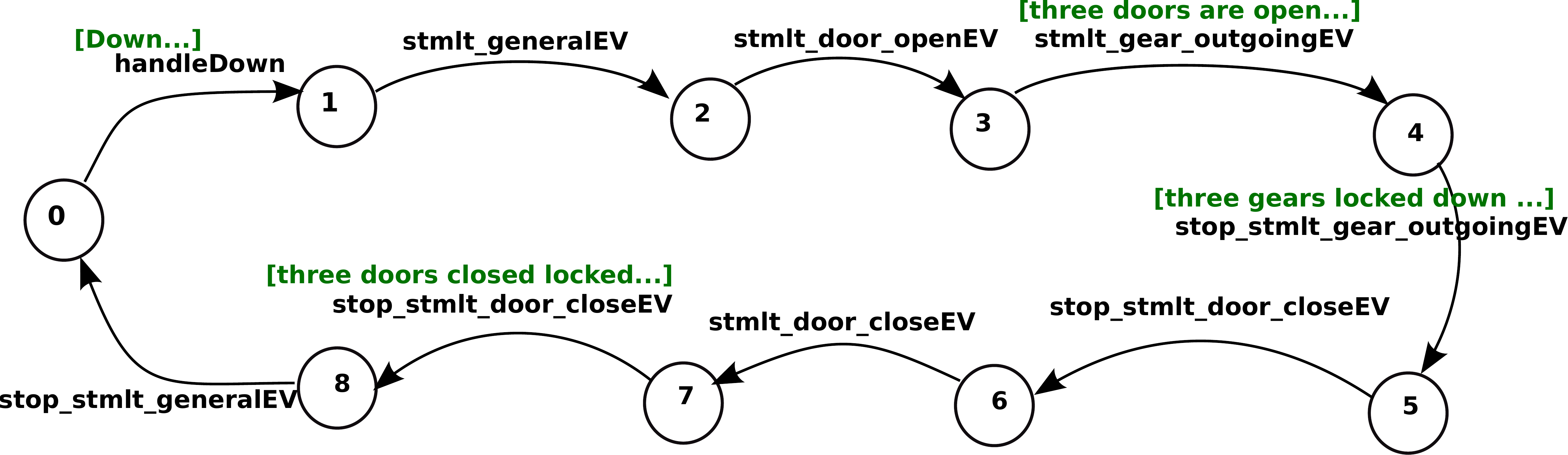}
\caption{The synthesis of the outgoing sequence}
\label{fig:outGoingSeq}
\end{figure}

The outgoing sequence is defined at the page 14 of the requirement
document.  It starts with the order DOWN and is finished when the
gears are extended and door closed. Between these events we have an
interaction involving the digital part which issues the orders, the
physical part which executes the orders and the sensors which provide
the various states of the gears and doors.
In Fig.~\ref{fig:outGoingSeq} we depict the overall behaviour of the
outgoing sequence, we can distinguish the main events and how they
interact with the other components which are modelled: sensors and
actuators. The labels of the transitions in the figure are the events
that model the behaviour of the digital part with respect to the
outgoing sequence. Note that the guards of the events depend on the
current state of the transition system and on the state of the digital
part (the sensors). In  Fig.~\ref{fig:outGoingSeq} we use the
brackets to indicate the elements of the guards.  The second sequence
---retraction sequence--- is modelled in the same way.

From the methodological point of view, we are still in the
\textbf{level~ 1} of Fig.~\ref{fig:principle}; several horizontal
refinements, depicted by the dashed arrows in the figure, are necessary
to integrate gradually the variables and the related events.
 
\subsubsection{Method of the construction of the events}
%CA Details :-(
\label{section:eventFamily}
The starting point is the state space obtained with the three
categories of variables at the interface of the digital part and the
fourth category of the internal variables. We define a family of
events related to each category of the interface variables:
\textit{monitoring events}, \textit{control events}, \textit{sensing
  events}.  Note that the construction of the abstract model is
achieved gradually by incorporating the variables and the events.

\paragraph{\textbf{Monitoring events family}}
The state output variables are used for monitoring; to set these
variables according to the current state of the controller and the
input data, we define for each variable (for example
\textit{gear\_locked\_down}) of this category an event named after the
variable (\textit{monitor\_gear\_locked\_down}) with the prefix
\textit{monitor\_}. The events of this family modify the appropriate
state output variables.  An example of these events is the following
one where the state output variable $gears\_locked\_down$,
is modified according to the
current state and the input variable $gear\_extended$. 

\begin{center}
\begin{boxedminipage}{12cm}
\begin{tabbing}
\=\hspace{0.5cm}\=\hspace{0.5cm}\=\hspace{0.3cm}\=\hspace{1cm}\kill
\> \textsf{event}   monitor\_gears\_locked\_Down  \\
\>  /* page 7 : the outputs are synthesised by each module \\
\>     ~~~~from sensors data and from the situation awareness \\
\>     page 15 : gear\_locked\_down = true iff the 3 gears are seen as locked \\
\>     ~~~~in extended position */ \\
\> \textsf{where} \\
\>\>     $ @g1~~ ran(gear\_extended)  = \{TRUE\} $ // the 3 gears are seen as locked \\
\>\>      @gano anomaly = FALSE // no anomaly detected \\
\> \textsf{then}\\
\>\>     $@a1~~  gears\_locked\_down := TRUE $\\
\>\>     $@a2~~  greenLight :=   lightON $\\
\>\>      $@a3~~  orangeLight :=  lightOFF $\\
\>\>      $@a4~~  redLight :=  lightOFF $\\
\> \textsf{end}
\end{tabbing}
\end{boxedminipage}
\end{center}

\paragraph{\textbf{Control events family}}
The order output variables are used to specify the actions that
stimulate the physical devices; for this purpose, each variable of
this category is set with an event named after the variable with the
prefix \textit{stmlt\_}, for example \textit{stmlt\_general\_EV,
  stmlt\_extend\_EV}, etc. These events use the internal variables,
the input variables and the output variables.  An example of these
events is the following one where the order output variable
$extend\_EV$ is modified.

\begin{center}
\begin{boxedminipage}{12cm}
\begin{tabbing}
\=\hspace{0.5cm}\=\hspace{0.5cm}\=\hspace{0.3cm}\=\hspace{1cm}\kill
\> \textsf{event}  stmlt\_gear\_outgoing  \\
 \>   /* stimulate gear outgoing electro valve \\
 \>    ~~~~  ** action 3 ** of outgoing sequence \\
 \>    ~~~~  once the three doors are in the open position */ \\
 \>     \textsf{where}\\
\>\>      $ @g0~~ general\_EV = TRUE$\\
 \>\>      $@g1~~ order = hDown$\\
 \>\>     $ @g2~~ ran(handle) = \{hDown\}$\\
 \>\>     $ @g3~~ ran(door\_closed) = \{FALSE\}$ // the three doors are in the open position\\
\>\>       $@g4~~ ran(door\_open) = \{TRUE\}$\\
\>\>       $@next~~ nextOGseq = 3$\\
\>\>       $@gano~~ anomaly = FALSE $// no anomaly detected\\
\>\>      $ @notretract~~ retract\_EV = FALSE$\\
 \>     \textsf{then}\\
\>\>      $ @a1~~ extend\_EV := TRUE$\\
\>\>      $ @a2~~ nextOGseq := nextOGseq + sequenceStep$~~// action 4 or action 2\\
 \> \textsf{end}
\end{tabbing}
\end{boxedminipage}
\end{center}

Associated with these events to stimulate the physical devices, we
have as many events to stop the stimulation of the devices.  These
events have their name prefixed with \textsf{stop\_}; an example is
\textsf{stop\_stmlt\_gear\_outgoing} which sets the variable $extend\_EV$ to
$FALSE$.

\smallskip 
\paragraph{\textbf{Sensing events family}} The input variables are
read by the digital part but they are set by the sensors which are
outside the digital part.  For these variables we define the events
named after the variables with the prefix \textit{sense\_}; some examples are
\textit{sense\_gear, sense\_door, ...}.  At this step of feature
augmentation, these events \textit{anticipate} their real future
specifications, which are related to the physical part introduced
latter.
An example of these events is the following;
 
\begin{center}
\begin{boxedminipage}{12cm}
\begin{tabbing}
\=\hspace{0.5cm}\=\hspace{0.5cm}\=\hspace{0.3cm}\=\hspace{1cm} \kill
\> \textsf{event}  sense\_gear // anticipated\\
\>    \textsf{where}\\
\>\>      $@noAno~~ anomaly = FALSE$ \\
\>    \textsf{then}\\
\>\>      $@a0~~ gear\_extended :∈ (TRIPLE×GEAR) → BOOL$\\
\> \textsf{end}
\end{tabbing}
\end{boxedminipage}
\end{center}

\medskip Note that these events are completed with specific variables
and events used to detect anomaly, to detect the failure of physical
devices, etc.

\subsubsection{Properties integrated in the abstract model} The
properties to be proved (requirements given in pages 18-19 of the
requirement document) are formalised as first order predicates
integrated into the invariant of the abstract model and proved along
the horizontal refinement. Most of the normal mode requirements are
safety properties. Requirement $R_1$ needs a specific treatment
presented in the sequel.

\begin{itemize}
\item Requirements $R_{21}$ and $R_{22}$.

\smallskip
\begin{tabular}{|l|p{11cm}|}
  \hline
  $R_{21}$ &  We can not observe a retraction sequence 
(consequence of the order $hUp$) if the handle is down. 
Using the enumerated set $HSTATE$ which permits only
 one value from two for the variable $order$, \\
  \hline
  \multicolumn{2}{|c|}{$order = hDown ⇒ ran(handle) ≠ \{hUp\}$}\\
  \hline
\end{tabular}

\medskip 
\begin{tabular}{|l|p{11cm}|}
\hline
$R_{22}$ & In a similar way we cannot observe an outgoing sequence 
(consequence of the order $hDown$) if the handle is up.\\
\hline
\multicolumn{2}{|c|}{$order = hUp ⇒ ran(handle) ≠ \{hDown\}$}\\
\hline
\end{tabular}

\medskip 
\item Requirements $R_{31}$ and $R_{32}$.

\smallskip
\begin{tabular}{|l|p{11cm}|}
\hline
$R_{31}$ & The gears outgoing event occurs if doors are open locked\\
\hline
\multicolumn{2}{|c|}{$(extend\_EV = TRUE ⇒ ran(door\_open) = \{TRUE\})$} \\
\hline
\end{tabular}

\medskip 
\begin{tabular}{|l|p{11cm}|}
\hline
$R_{32}$ & The gears retraction event occurs if doors are open locked\\
\hline
\multicolumn{2}{|c|}{$(retract\_EV = TRUE ⇒ ran(door\_open) = \{TRUE\})$} \\
\hline
\end{tabular}

\medskip 
\item Requirement $R_{41}$ and $R_{42}$.

\smallskip
%  @PropReq41
\begin{tabular}{|l|p{11cm}|}
\hline
$R_{41}$ &  Opening and closing doors electro-valve are not 
stimulated simultaneously\\
\hline
\multicolumn{2}{|c|}{$ ¬(open\_EV = TRUE ∧ close\_EV = TRUE)$}\\
\hline
\end{tabular}

\medskip 
%  @PropReq42
\begin{tabular}{|l|p{11cm}|}
\hline
$R_{42}$ &   Outgoing and retraction gears electro-valve are
 not stimulated simultaneously\\
\hline
\multicolumn{2}{|c|}{$ ¬(extend\_EV = TRUE ∧ retract\_EV = TRUE)$}\\
\hline
\end{tabular}

\medskip 
\item Requirement $R_5$.

\smallskip
%  @PropReq51
\begin{tabular}{|l|p{11cm}|}
\hline
$R_{51}$ & It is not possible to stimulate the manoeuvring 
 EV (opening, closure, outgoing or retraction) without 
stimulating the general EV  \\
\hline
\multicolumn{2}{|c|}{$((open\_EV = TRUE ∨ close\_EV = TRUE$}\\
\multicolumn{2}{|c|}{$~~~ ∨ extend\_EV = TRUE ∨ retract\_EV = TRUE)$}\\
\multicolumn{2}{|c|}{$~~~ ⇒ general\_EV = TRUE)$} \\
\hline
\end{tabular}

\end{itemize}

%%%%%%%%%%%%%%%%%%%%%%%%%%%%%%%%%%%%%%%%%%%%%%%%%%%%%%%%%%%%%%%%%%%%%%%
%%%%%%%%%%%%%%%%%%%%%%%%%%%%%%%%%%%%%%%%%%%%%%%%%%%%%%%%%%%%%%%%%%%%%%%
%%%%%%%%%%%%%%%%%%%%%%%%%%%%%%%%%%%%%%%%%%%%%%%%%%%%%%%%%%%%%%%%%%%%%%%

\subsection{Capturing the behaviours of the physical part}
\begin{figure}[htbp]
\centering
\includegraphics[width=\linewidth]{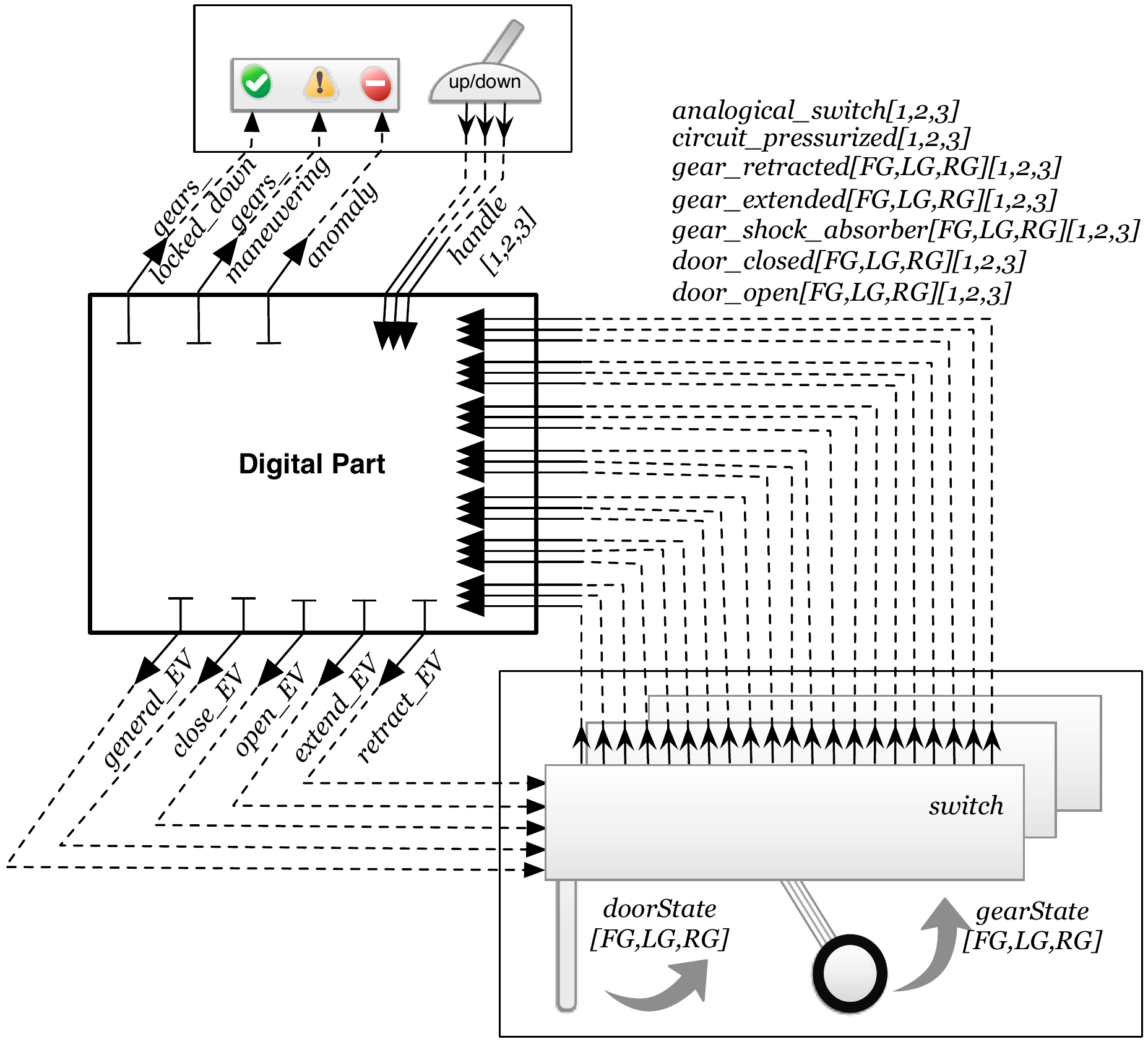}
\caption{The digital part connected to the physical environment}
\label{fig:digitalpart+physical}
\end{figure}

One more step of the refinement by feature augmentation is the
addition of the behaviour of physical devices: the sensors, the doors,
the gears, as illustrated by Fig.~\ref{fig:digitalpart+physical}.
Several refinement steps are necessary to integrate gradually the
variables and events related to these devices; this corresponds
to the dashed line in the \textbf{level 2} of Fig.~\ref{fig:principle}.

\subsubsection{Door behaviour}
\begin{figure}[htbp]
\centering
\includegraphics[width=.65\linewidth]{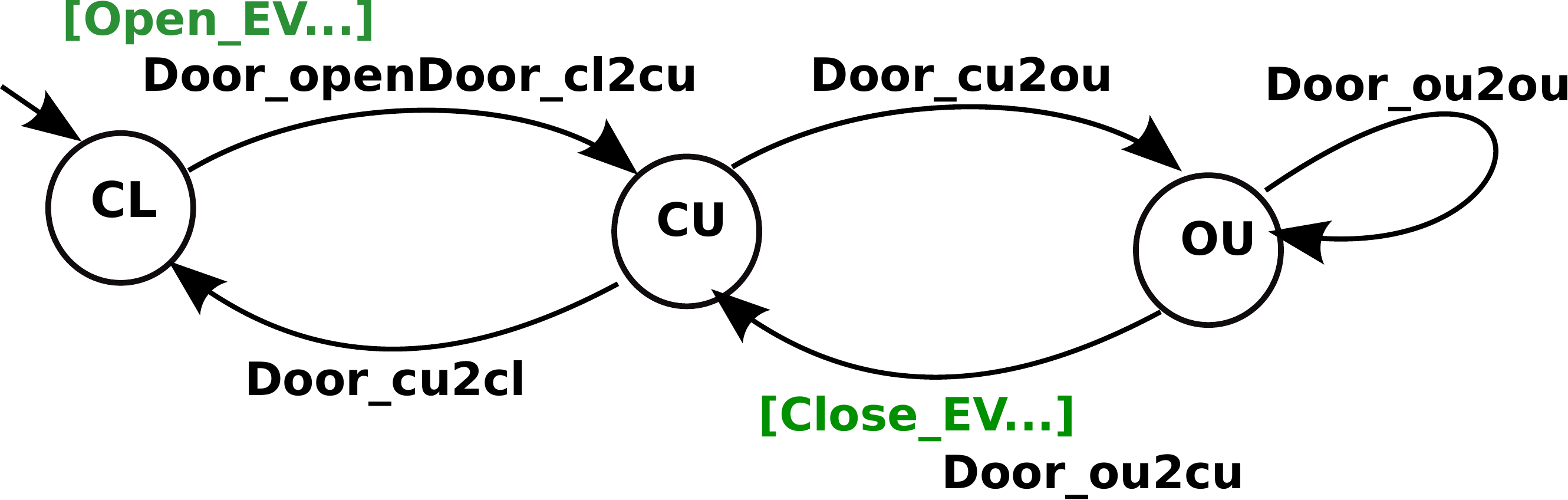}
\caption{Behaviour of doors}
\label{fig:doorBehaviour}
\end{figure}

The door behaviour is described at page 11 of the requirement
document.  It is first captured with a state automata; the transitions
of the automata are then described as events. For this purpose we use
a transition function $doorState ∈ DOOR → DSTATE$ where $DSTATE =
\{ClosedLocked, ClosedUnlocked, OpenUnlocked\}$ is the enumerated set
of the identified door states (see Fig.~\ref{fig:doorBehaviour}). The set $DOOR$
contains the three doors. The function $doorState$ is a total
function; this captures the requirement that all the three doors are
controlled via the state transition. 

The starting transition of the door behaviour is enabled by the
$open\_EV$ order\footnote{In Fig.~\ref{fig:doorBehaviour} the brackets
  indicate the event that contributes to enable the transition.} given
by the digital part. Therefore there is a synchronisation between the
digital part and the doors.  The remaining transitions are handled
with events which are conveniently guarded in such a way that we have
the correct ordering of the doors. Note that the three doors are
simultaneously controlled by the orders issued by the digital part.
We give below the description of the starting event (\texttt{Door\_openDoor\_cl2cu}).

\begin{center}
\begin{boxedminipage}{12cm}
\begin{tabbing}
\=\hspace{0.5cm}\=\hspace{0.5cm}\=\hspace{0.3cm}\=\hspace{1cm}\kill
\> \textsf{event} Door\_openDoor\_cl2cu \\
\>  /* Door's Behaviour (for the three doors)\\
\>   ~~~  The first transition of the Door automata\\
\>   ~~~  enabled when the action open\_EV is performed by the control system */\\
\>    \textsf{where}\\
\>\>      $@g1~~ open\_EV = TRUE$ // all doors EV are on \\
\>\>      $@g2~~ ran(doorState) = \{notOpenLocked\}$ \\
\>    \textsf{then} \\
\>\>      $@a1 doorState:= DOOR×\{notOpenNotLocked\}$ // door is being opened\\
\> \textsf{end}
\end{tabbing}
\end{boxedminipage}
\end{center}

In the same way the transition starting from the open to
close position is synchronised with the $close\_EV$ order given by the
digital part. Therefore we have the complete interaction between
between the orders from the digital part and the doors.
 
\subsubsection{Gear behaviour}
\begin{figure}[htbp]
\centering
\includegraphics[width=.65\linewidth]{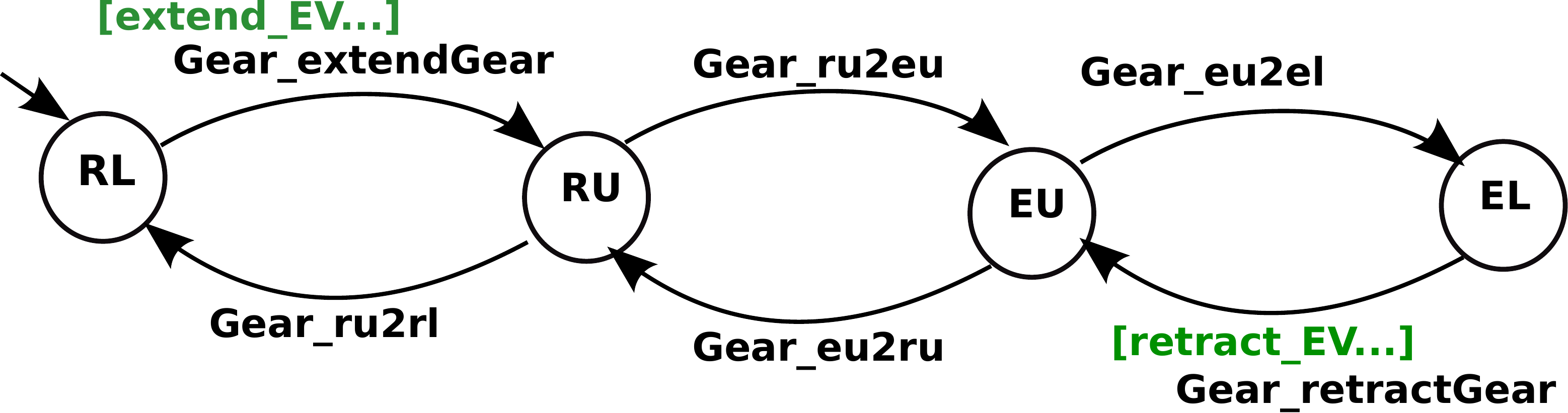}
\caption{Behaviour of the gears}
\label{fig:gearBehaviour}
\end{figure}

The gear behaviour is specified in the same way as the doors, see Fig.~\ref{fig:gearBehaviour}. A
specific transition function is used: $gearState ∈ GEAR → GSTATE$
where $GSTATE = \{RetractedLocked, RetractedUnlocked,
ExtendedUnlocked, ExtendedLocked\}$ is an enumerated set of gear
states. The labels of the transition correspond to the events that
model the behaviour of the gears. The set $GEAR$ contains the three
gears.

\subsubsection{Sensor behaviour}
The sensors behave like the observers of the door states and the gear
states. When a door reaches the \textsf{open} state, then the related
sensor events set the variable $door\_open$ to $TRUE$.

The related events in the specification are prefixed with
\textit{sense\_}.  The modelling of the sensors follows faithfully the
requirement document; that is each sensor is made of three
micro-sensors; each of them has its own state.  We illustrate the
modelling with the following event which deals with the extension of
one micro-sensor\footnote{identified by 1} of the front gear; the
event sets the corresponding value in the interface variable
($gear\_extended$).

\begin{center}
\begin{boxedminipage}{12cm}
\begin{tabbing}
\=\hspace{0.5cm}\=\hspace{0.5cm}\=\hspace{0.3cm}\=\hspace{1cm}\kill
\> \textsf{event}  sense1\_FGE\_OK // is sensor1 of Front Gear extended ? \\
 \> \textsf{refines} sense\_gear\\
\>    \textsf{any} nge\\
\>    \textsf{where}\\
\>\>      $@noAno~~ anomaly = FALSE$ \\
 \>\>     $@g1~~ gearState(FG) = LockE$ // the Front Gear is seen Locked Extended\\
 \>\>     $@g2~~ nge ∈ (TRIPLE×GEAR) → BOOL$\\
 \>\>     $@g3~~ nge = gear\_extended <\!\!+  \{  (1 \mapsto FG) \mapsto  TRUE\}$ // update of the var\\
\>    \textsf{then}\\
\>\>      $@ea2~~gear\_extended := nge$ // that is (1 $\mapsto$ FG) := TRUE \\
\>  \textsf{end}\\
\> \textsf{end}
\end{tabbing}
\end{boxedminipage}
\end{center}

Note that a sensor can have an abnormal functioning.  To
simulate this anomaly on the sensor, the corresponding event could set
the state of the gear to the wrong value. For this purpose, we
define in our model such events, named likely
\textsf{sense1\_FGE\_KO}, which set a wrong value. 
%Style :-( take that into account.

%%%%%%%%%%%%%%%%%%%%%%%%%%%%%%%%%%%%%%%%%%%%%%%%%%%%%%%%%%%%%%%%%%%%%%%
%%%%%%%%%%%%%%%%%%%%%%%%%%%%%%%%%%%%%%%%%%%%%%%%%%%%%%%%%%%%%%%%%%%%%%%
%%%%%%%%%%%%%%%%%%%%%%%%%%%%%%%%%%%%%%%%%%%%%%%%%%%%%%%%%%%%%%%%%%%%%%%

\subsection{Handling some reachability requirements}

%\subsubsection{Introducing the reachability property (requirement 2)}

Based on the idea of Lamport's logical
clocks~\cite{DBLP:journals/cacm/Lamport78}, we implement a technique
that captures the reachability requirement $R_1$ given in page 13 of
the requirement document.  For that purpose, we introduce the notion
of \textit{control cycle}.  A \textit{control cycle} is a period of
time during which one can observe several events, especially a chain
of events denoting an outgoing sequence or a retraction sequence; a
typical control cycle is one starting with an event which denotes the
$hDown$ order and terminating by an event which denotes the fact that
``\textit{the gears are locked down and the doors are seen closed}'';
similarly, another control cycle is started when the handle triggers
an order $hUp$.  A dedicated variable $endCycle$ is used to control
the start and the end of each control cycle.

\begin{figure}[htbp]
\centering
\includegraphics[width=.6\linewidth]{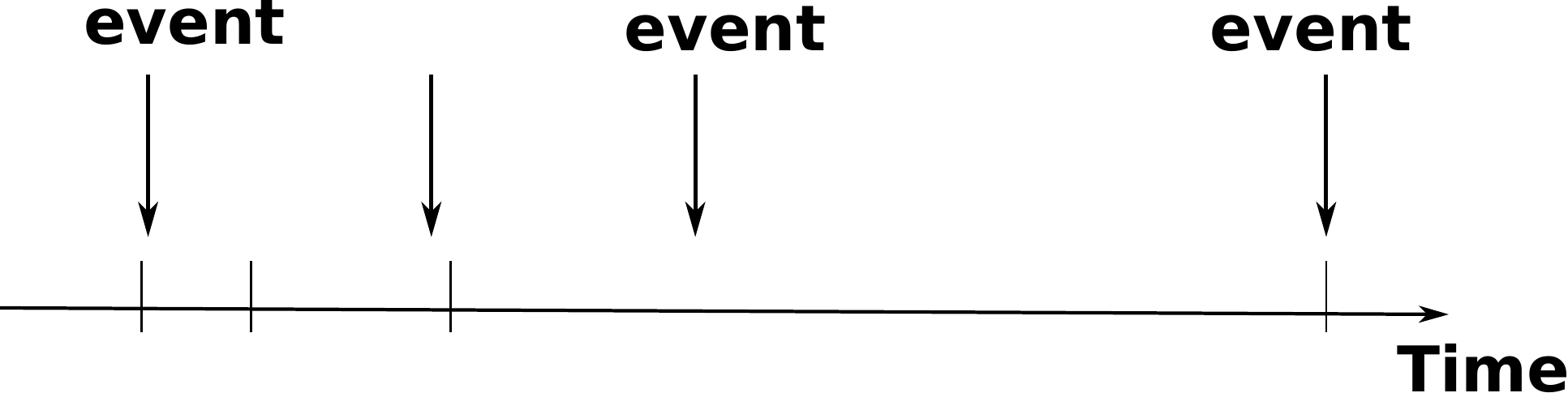}
\caption{Events and timestamps}
\label{fig:eventTimestamps}
\end{figure}

Assume that we have observable events that occur along the time and
that denote our events of interest\footnote{These events are not to be
  confused with Event-B events.}; for instance the starting of an
outgoing sequence, a door closed, a gear locked in a position,
etc. Each such event can be stamped with the timestamp of its
occurrence, thus if we have the set of observed events we can define
at least a partial ordering of these events (see
Fig.~\ref{fig:eventTimestamps}).

Given a set $obsEvents$ of events and a logical clock modelled as a
natural number, the occurrences of the events can be ordered by the
timestamp given by the clock. In our case two events cannot happen at
the same time. We use a partial function $ldate ∈ obsEvents \pfun
\nat$ to record the timestamps of the events.  We can compare and
reason on the timestamps of any events happening during a sequence and
specifically within the specific event sequence called \textit{control
  cycle}.

For example, in the normal mode, we observe the event 
``\textit{the door is closed and the gear extended}'' (named
\textsf{dcge}) at the end of a cycle, if the event ``\textit{order DOWN is
  given}'' (named \textsf{downH}) occurs and is maintained (no event
\textsf{upH} occurs). If these events have respectively the specific
timestamps $dj$ and $di$, then we can compare $di$ and $dj$ and
also examine the events which happen between $di$ and
$dj$. Accordingly the property $R_{1bis}$ of the requirement is expressed
as follows: 

\begin{center}
\begin{boxedminipage}{12cm}
\begin{tabbing}
\=\hspace{0.5cm}\=\hspace{1cm}\=\hspace{0.3cm}\=\hspace{1cm}\kill
$∀dj.( ((dj ∈ \nat) ∧ (dcge ∈ dom(ldate)) ∧ (dj = ldate(dcge))$\\
\> $~~~~~~~~~~~~~~ ∧ (endCycle = TRUE) ∧ dj < llc) ⇒ $\\
\>\> $∃di.( (dd∈ \nat) ∧ (downH ∈ dom(ldate)) ∧ (di = ldate(downH)) ∧ (di < dj) ∧$\\
\>\>\>        $∀ii.(ii∈ \nat ∧ di ≤ ii ∧ ii < dj ⇒ ldate∼[\{ii\}] ≠ \{upH\})))$
\end{tabbing}
\end{boxedminipage}
\end{center}

The above property means that if we reach the end of a control cycle
where the door is closed and the gear extended at a given timestamp
($dj$), then we should have an order $hDown$ issued at a timestamp
$di$ less that $dj$ and maintained between $dj$ and $di$; the outgoing
sequence is not interrupted by an order $hUp$ which would start
another cycle.  Consequently we have expressed the property
$R_{11bis}$.  Property $R_{12bis}$ can be expressed in a similar
manner.

To put in practice in Event-B with Rodin, we defined the set
$obsEvents$ in the context of our machines, and the above property is
included in the invariant of the abstract model.

This step of the horizontal refinement process is the \textbf{level~
  3} of Fig.~\ref{fig:principle}.

\subsection{Well-ordering the control of the physical part}

The modelled behaviour of the landing system especially the model of
the digital part should attend the two main objectives given at pages
13 of the requirement document: \textit{"to control the physical
  devices"} and \textit{"to monitor the system and inform the pilot"}.

Control systems have essentially three main steps: sensing of input
variables (set by external mechanisms), making a decision (compute
some values with respect to the current state), actuating the
controlled mechanisms (by modifying their values via the
actuators). This constitutes the elementary interaction cycle
\textsf{sense/decision/order} which is repeatedly applied to control a
system.

The behaviour described by our model should be conform to the
interaction cycle.  It is the case; each control cycle initiated by
the pilot via the handle, to achieve an outgoing sequence or a
retraction sequence, is made of several interaction cycles
\textsf{sense/decision/order}.  Indeed, in the normal mode of the
landing system, the outgoing sequence and the retraction sequence
structure the first objective of controlling the physical device; a
control cycle is performed according to each sequence.  As explained
in Sect.\ref{section:stepwiseAbsM} (page \pageref{page:outgoingSeq})
our model is structured according to the two sequences; each sequence
is made of a chain of events which guards are defined in such a way
that\footnote{they follow the state transition} the control implements
the interaction cycle \textsf{sense/decision/order} which involve the
events of the three families identified (please, see page
\pageref{section:eventFamily}).  Note that the \textsf{sense} step
involves the events of the \textit{sensing} family; the
\textsf{decision} step involves the events of the \textit{monitoring}
family; the \textsf{order} step involves the events of the
\textit{control} family. Moreover the events of the
\textit{monitoring} family achieve the second objective of the landing
system.

%====================================================

\section{Building the Digital Part with Structural Refinement}
% raffinement vers le concret et prise en compte de nouvelles propriétés
\label{section:refinedModel}
% -*- coding: utf-8 -*-
%
%-----------------------------
% Landing System ABZ'2014
% 
%\section{Structural Refinement}
% raffinement vers le concret et prise en compte de nouvelles propriétés

Structural refinement starts when we have finished the construction of
the abstract model of our system integrating the digital part, the
gear boxes and the pilot interface, and we have proved all the
necessary requirements.

\begin{figure}[htbp]
	\centering
	\includegraphics[width=\linewidth]{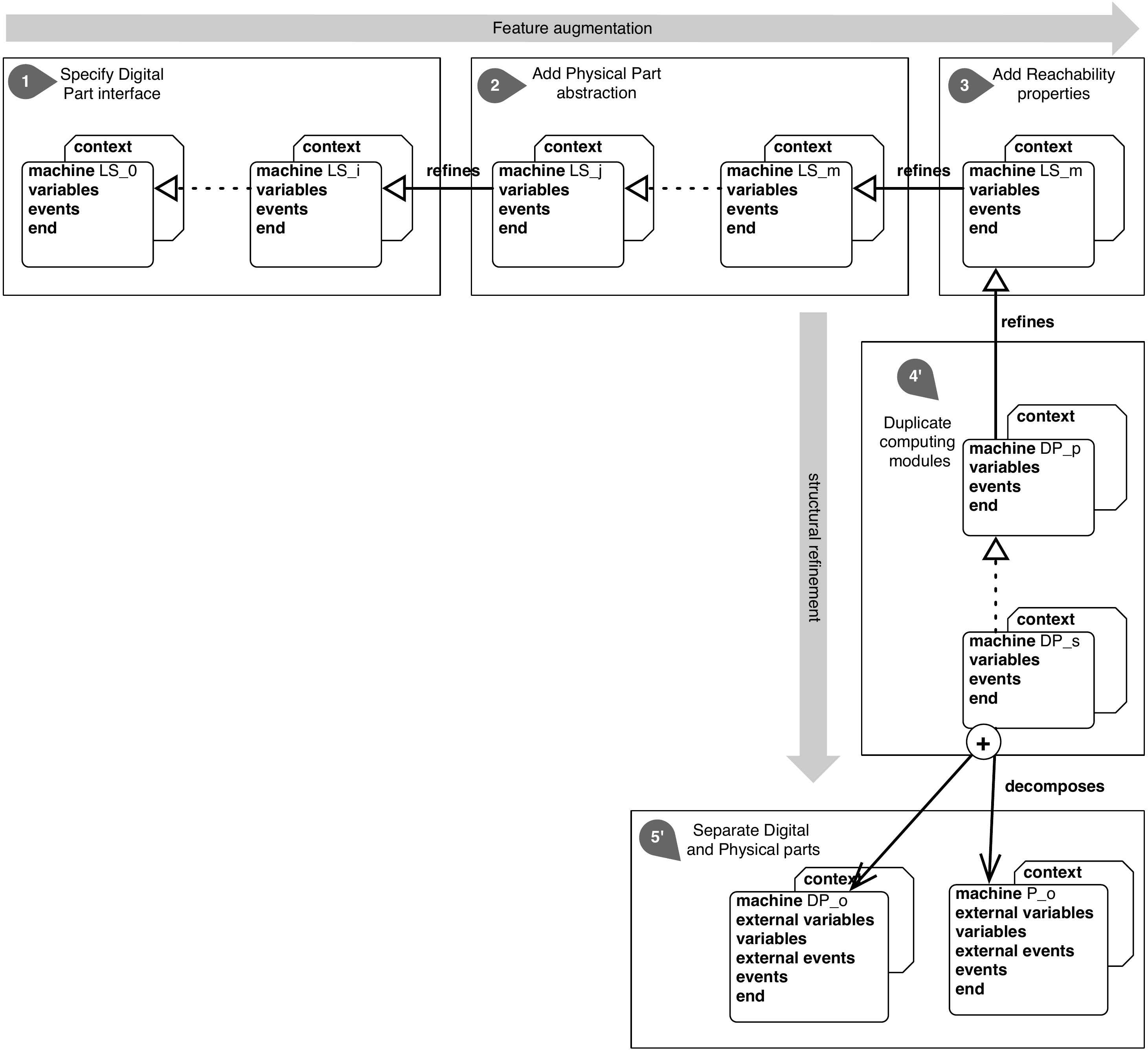}%width
        \caption{The adopted modelling approach (a variant of Fig.\ref{fig:principle}) }
	\label{fig:principle:variant}
\end{figure}

During the structural refinement only the digital part will be refined
with the objective to build the software system. Consequently we
concentrate the refinement on the variables and events of the digital
part. The variables and events which are specific to the behaviour of
the physical part are not refined but we keep them in the model in
order to preserve \textit{animation capabilities}. This approach is
very pragmatic.  In addition to the proofs, the animation provides a
concrete view of the system behaviour and this helps in gaining
confidence in the model and also in a validation process with a client
as example.  Therefore the model is not decomposed as we have
presented in Sect.~\ref{section:reqAnalysis}: the model will be
decomposed into two models at the end of the structural refinement, as
depicted in Fig.~\ref{fig:principle:variant}.

In the following, we explain how we refine the digital part; the
starting point is the global abstract model obtained from the previous
section (Sect.~\ref{section:abstratModel}).

%%%%%%%%%%%%%%%%%%%%%%%%%%%%%%%%%%%%%%%%%%%%%%%%%%%%%%%%%%%%%%%%%%%%%%%
%%%%%%%%%%%%%%%%%%%%%%%%%%%%%%%%%%%%%%%%%%%%%%%%%%%%%%%%%%%%%%%%%%%%%%%
%%%%%%%%%%%%%%%%%%%%%%%%%%%%%%%%%%%%%%%%%%%%%%%%%%%%%%%%%%%%%%%%%%%%%%%

\subsection{Refinement of the digital part}
\label{section:refineDigital}
% -*- coding: utf-8 -*-
%
%-----------------------------
% Landing System ABZ'2014
% 
% \subsection{Refining the digital part}
%
%

The requirement document details the inner structure of the digital
part, which is made of two redundant computing modules (see the
Digital architecture depicted in Fig.5 of the requirement
document). The policies to compose the inputs and the outputs of the
two modules are explained (pages 5-6 of the requirement document).
Some structural refinements are helpful to capture faithfully this
architecture of the digital part.

\subsubsection{Introducing the two computing modules with a refinement}
Both modules have the same interface (input and output variables)
inherited from the abstract model of the digital part.

\begin{figure}[htbp]
\centering
\includegraphics[width=\linewidth]{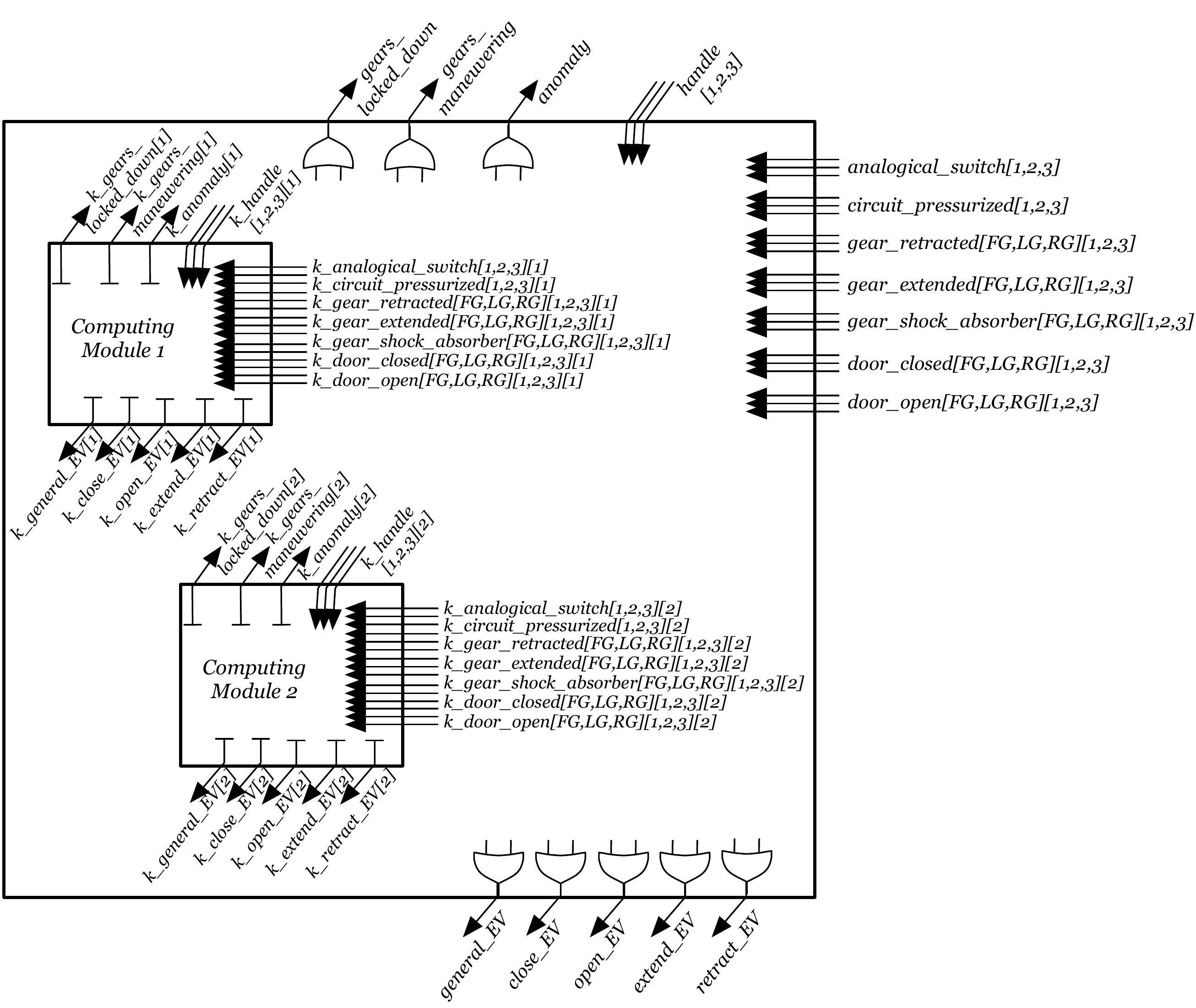}
\caption{Principle for adding the two modules}
\label{figure:kmodule}
\end{figure}

As far as the method is concerned, the principle of our solution is depicted in Fig.~\ref{figure:kmodule}.
  Each interface variable of a module
$k$ (where $k \in \{1,2\}$) is inherited from a variable (for instance
$gear\_extended$) of the digital part of the abstract model and it is
denoted by $k\_gear\_extended(k)$ where $(k)$ is an index. The prefix
$k\_$ of the variable name enables us to keep the same name but to
differentiate the abstract variable name and the refined one. An
enumerated set $CompModule = \{1,2\}$ is used for the indexes.
Therefore each interface variable of the computing modules is
specified with the following shape, where the abstract variables are
indexed.

\begin{center}
\begin{boxedminipage}{12cm}
$k\_handle ∈ CompModules → (TRIPLE→ HSTATE)$\\
$k\_gear\_extended ∈ CompModule  → ((TRIPLE×GEAR) → BOOL)$\\
$k\_gear\_retracted ∈ CompModule  → (TRIPLE×GEAR) → BOOL$\\
$k\_analogical\_switch ∈ CompModule  → (TRIPLE → AnalSWSTATE)$\\
$k\_door\_closed ∈ CompModule  → (TRIPLE × DOOR) → BOOL$\\
$k\_door\_open ∈ CompModule  → (TRIPLE × DOOR) → BOOL$\\
$\cdots$\\
$k\_general\_EV ∈ CompModule  → BOOL$\\
$k\_extend\_EV ∈ CompModules → BOOL$\\
$k\_close\_EV ∈ CompModule  → BOOL$\\
$k\_open\_EV ∈ CompModule  → BOOL$\\
$\cdots$
\end{boxedminipage}
\end{center}

Notice, at this stage, the interface of the two modules are not linked
with that of the abstract module.  This is achieved via further
refinement features: the binding between the variables and their
k-indexed forms. 
%  Therefore, for each variable of the digital part
% which is also present at the computing module interface with the
% k-indexed form, a binding invariant should be given.

\subsubsection*{Spawning the inputs  inside the computing modules}

We have to specify that the inputs of the digital part (either from
the cockpit or from the sensors) are the same ones for the computing
modules. The principle adopted here is that the value of each input
variable (for example $handle$) at the abstract level is pushed in the
corresponding variable (for example $k\_handle$) of each computing
module.  As the inputs of the modules are (and should be) the same, an
invariant is defined in each case of variable spawning in order to
guarantee the correctness of the binding between the input variable of
the digital part and the same input of the computing modules.

\begin{center}
\begin{boxedminipage}{12cm}
$handle ∈ ran(k\_handle)$ ~~~~~ ~~~ \rm /* binding invariant */ \\
$gear\_extended ∈ ran(k\_gear\_extended)$ ~~~~~ ~~~ \rm /* binding invariant */
\end{boxedminipage}
\end{center} 

We use the following event pattern to spawn the variables at the
interfaces of the computing modules.

\begin{center}
\begin{boxedminipage}{12cm}
\begin{tabbing}
\=\hspace{0.5cm}\=\hspace{0.5cm}\=\hspace{0.3cm}\=\hspace{1cm}\kill
\> \textsf{event} spawn\_handleDown // spawn handleDown within the k CompModules\\
\>  \textsf{where} \\
\>\>   @g1~~ $ran(handle) = \{hDown\}$\\
\>    \textsf{then} \\
\>\>  @a1~~ $k\_handle := \{1 \mapsto (TRIPLE×(ran(handle))), $\\
\>\> $~~~~~~~~~~~~~~2 \mapsto (TRIPLE×(ran(handle)))\}$\\
\>  \textsf{end}
\end{tabbing}
\end{boxedminipage}
\end{center}

Consequently one identified \textbf{specification rule} is that a new event is
introduced along with each new k-indexed variable. This event should
copy the variable at high level (the digital part) into the indexed
variables at the low level. This specification rule is reusable in all
such cases where variables from a module should be buried inside
submodules.
 
Furthermore, the existing events, which guards or actions involve 
spawned variables, should be refined by extending their guards and
actions in order to satisfy the binding between the variables and the
associated k-indexed variables.

As far as the input from the sensors are concerned, specific treatment
is needed in order to take account of the separate behaviours of the
three micro-sensors that can send different values event wrong
values. For this purpose, if we consider each (front, left, right) door or
gear,  two events are needed to specify the impact of the
micro-sensor behaviour on the k-indexed input variables of the
computing modules; one event gives the right value of the micro-sensor, whereas
the other event gives the wrong value.

One noticeable feature in this case is that when we have a
nondeterministic event of abstract level (as for the value of the
sensors), then in the refinement the event should be refined (not
extended). This is a second identified \textbf{specification rule}.

As an example, we consider the sensor 2 from the three sensors that
sense if the left door is open ($LDO$), hence the event names
$sense2\_LDO\_OK$, $sense2\_LDO\_KO$.

\begin{center}
\begin{boxedminipage}{12cm}
\begin{tabbing}
\=\hspace{0.5cm}\=\hspace{0.5cm}\=\hspace{0.3cm}\=\hspace{1cm}\kill
\> \textsf{event}  event sense2\_LDO\_OK // seonsor2 of LDoor\\
\>  \textsf{extends} sense2\_LDO\_OK\\
\>     \textsf{any} nkdo\\
 \>    \textsf{where}
\>\>       $@noano~~ ∀kk.(kk ∈ CompModules ⇒ (k\_anomaly(kk)  = FALSE))$\\
\>\>       $@dnkge~~ nkdo ∈ CompModules → ((TRIPLE×DOOR) → BOOL)$\\
\>\>       $@nge~~ nkdo~ = CompModules × \{door\_open <+ \{(2 \mapsto LD)\mapsto TRUE\}\}$\\
\>   \textsf{then}\\
\>\>      $@ea1 k\_door\_open := nkdo$\\
\>  \textsf{end}
\end{tabbing}
\end{boxedminipage}
\end{center}

\begin{center}
\begin{boxedminipage}{12cm}
\begin{tabbing}
\=\hspace{0.5cm}\=\hspace{0.5cm}\=\hspace{0.3cm}\=\hspace{1cm}\kill
\> \textsf{event}   event sense2\_LDO\_KO // sensor2 of FDoor (simulating malfunctioning)\\
\>  \textsf{extends} sense2\_LDO\_KO\\
\>     \textsf{any} nkdo\\
\>    \textsf{ where}\\
 \>\>      $@noano~~ ∀kk.(kk ∈ CompModules ⇒ (k\_anomaly(kk)  = FALSE))$\\
 \>\>      $@dnkge~~ nkdo ∈ CompModules → ((TRIPLE×DOOR) → BOOL)$\\
\>\>       $@nge~~ nkdo = CompModules × \{door\_open <+ \{(2\mapsto  LD)  \mapsto FALSE\}\}$\\
\>   \textsf{ then}\\
\>\>       $@ea1~~ k\_door\_open := nkdo$\\
\>  \textsf{end}
\end{tabbing}
\end{boxedminipage}
\end{center}

\subsubsection{Merging the outputs of the computing modules}

The specification principle is as follows. As presented in the
requirement and depicted in Fig.~\ref{figure:kmodule}, the k-indexed
output variables (for example $k\_extend\_EV(1)$ and
$k\_extend\_EV(2)$) are merged using a logical OR to set the
corresponding variable (for example $extend\_EV$) at the output of the
digital part. Therefore the event that sets the variable should be
guarded by the availability of the merged value. As explained before,
a binding invariant should be provided for each variable and the
related k-indexed variable.

\begin{center}
\begin{boxedminipage}{12cm}
\begin{tabbing}
\=\hspace{0.5cm}\=\hspace{0.5cm}\=\hspace{0.3cm}\=\hspace{1cm}\kill
\> $k\_extend\_EV ∈ CompModules → BOOL$\\ 
\> $extend\_EV ∈ ran(k\_extend\_EV) $ /* binding invariant */ 
%$ k\_gears\_locked\_down ∈ CompModules → BOOL$
\end{tabbing}
\end{boxedminipage}
\end{center}
\smallskip

The merging is illustrated with the following event
$merge\_stmlt\_gear\_outgoing$, which updates the output variable
$extend\_EV$. Only the guard should be extended in the refinement
since the abstract event already has the right substitution to set the
variable.

\begin{center}
\begin{boxedminipage}{12cm}
\begin{tabbing}
\=\hspace{0.5cm}\=\hspace{0.5cm}\=\hspace{0.3cm}\=\hspace{1cm}\kill
\> \textsf{event} merge\_stmlt\_gear\_outgoing\\
\>  /* MERGING the result of the k module to stimulate the gear\_outgoing\\
\>\>     ** action 3 ** of outgoing sequence */ \\
\>  \textsf{extends} stmlt\_gear\_outgoing\\
\>\>    \textsf{any} kk \\
\>\>    \textsf{where} \\
\>\>     $ @gkk~~ kk ∈ CompModules $\\
\>\>    $ @theOR~~ k\_extend\_EV(kk) = TRUE $ /* OR : one of them is TRUE */ \\
\>  \textsf{end}
\end{tabbing}
\end{boxedminipage}
\end{center}
\smallskip

We identified this specification as a \textbf{promotion pattern} which
promotes the outputs of the modules at the level of the digital
part. This pattern is reusable in all cases of modules encapsulation.

As for the orders to the physical part, the output variables for the
cockpit result from the merging of the related output variables of the
computing module; therefore the event that modify these variables are
refined by extension of their guards. This is illustrated with the
following event.

\begin{center}
\begin{boxedminipage}{12cm}
\begin{tabbing}
\=\hspace{0.5cm}\=\hspace{0.5cm}\=\hspace{0.3cm}\=\hspace{1cm}\kill
\> \textsf{event}   merge\_monitor\_gears\_locked\_Down \\
\>  /* page 7 : the outputs are synthesised by each module \\
\>     from sensors data and from the situation awareness ... */\\
\>   \textsf{extends} monitor\_gears\_locked\_Down \\
 \>\>    \textsf{any} kk \\
\>\>     \textsf{where} \\
\>\>      $@gkk~~ kk ∈ CompModules$ \\
\>\>      $@gr1~~ k\_gears\_locked\_down(kk) = TRUE$ /* OR : meaning one is true */\\
\>  \textsf{end}
\end{tabbing}
\end{boxedminipage}
\end{center}

\subsubsection{Specifying the behaviour of the computing modules}

The two computing modules have the same behaviour. This behaviour is
the one sketched with the three categories of variables at the
interface of the digital part. Typically we have the events that
monitor the system and set the state output variables, the events that
give orders to the physical part and the impact of the input variables
on the state of the digital part.

The principle here is to define as for the variables, the k-indexed
form of the events related to the three categories of the interface
variables and the internal variables.
The following B event illustrates such an event of the computing modules. 
It is an anticipated event which will should be further refined.

\begin{center}
\begin{boxedminipage}{12cm}
\begin{tabbing}
\=\hspace{0.5cm}\=\hspace{0.5cm}\=\hspace{0.3cm}\=\hspace{1cm}\kill
\> \textsf{event} k\_stmlt\_general\_EV \\
\>  /* anticipated, will be refined  inside the k=1,2 modules */ \\
\>     \textsf{any} nkge \\
\>     \textsf{where} \\
\>\>      $@dkge~~ nkge ∈ CompModules → BOOL $\\
\>\>      $@vnke~~ nkge = CompModules × \{general\_EV\}$\\
\>     \textsf{then} \\
\>\>      $@a1~~ k\_general\_EV :=  nkge $\\
\>  \textsf{end}
\end{tabbing}
\end{boxedminipage}
\end{center}

%===================================================

%%%%%%%%%%%%%%%%%%%%%%%%%%%%%%%%%%%%%%%%%%%%%%%%%%%%%%%%%%%%%%%%%%%%%%%
%%%%%%%%%%%%%%%%%%%%%%%%%%%%%%%%%%%%%%%%%%%%%%%%%%%%%%%%%%%%%%%%%%%%%%%
%%%%%%%%%%%%%%%%%%%%%%%%%%%%%%%%%%%%%%%%%%%%%%%%%%%%%%%%%%%%%%%%%%%%%%%

\subsection{Decomposition into a digital part and a physical part}
\label{section:decomposeDigPhys}

A decomposition paradigm is supported by the Event-B method. Two
approaches exist\footnote{implemented in the Deploy Project.} for this
purpose: the Abrial'style decomposition (called the A-style
decomposition)~\cite{DBLP:journals/fuin/AbrialH07} based on shared
variables, and the Butler'style decomposition (called the B-style
decomposition)~\cite{DBLP:conf/ifm/Butler09,DBLP:conf/fmco/SilvaB10}
based on shared events.  In the A-style decomposition, events are
first split between Event-B sub-components and then shared variables
of the sub-components are used to introduce external events in the
sub-components; these external events should be refined in the same
way.  In the B-style decomposition, variables are first partitioned
between the sub-components and then shared events (which use the
variables of both sub-components) are split between the sub-components
according to the used variables.

We have used the A-style decomposition which is more relevant when
considering the events that describe the behaviour of two different
parts of the landing system. It is straightforward to list the events
that describe the behaviour of the physical part in order to separate
them from the events closely related to the digital part.  For this
purpose we have experimented the decomposition plugins of the Rodin
toolkit using the A-Style decomposition approach.

As shown in Fig.~\ref{fig:principle:variant}, the global abstract
model resulting from the horizontal refinement (\textbf{Level 3},
Fig.~\ref{fig:principle:variant}) is refined vertically (\textbf{Level
  4'}, Fig.~\ref{fig:principle:variant}).  The resulting refined model
should be decomposed into a digital part and a physical part
(\textbf{Level 5'}, Fig.~\ref{fig:principle:variant}).

Consequently, the digital part must be separated from the environment,
i.e. the physical part and the pilot interface.  The methodological
guide to achieve the decomposition is as follows:

\begin{itemize}

\item the digital part is made with the events defined in two families
  of events (see Sect.\ref{section:stepwiseAbsM}) related to the
  interface variables: \textit{monitoring events} and \textit{control
    events}.  
  \begin{itemize}
  \item The events in the monitoring family are all those with the
    names prefixed by \textit{monitor\_}. Examples are:
    \textsf{monitor\_gears\_locked\_Down, monitor\_gears\_maneu\-vering,
      monitor\_anomaly}.
  \item The events in the control events family are those with the
    names prefixed with \textit{stmlt\_} and
    \textit{stop\_stmlt\_}. Examples are \textsf{stmlt\_general\_EV},
    \textsf{stmlt\_door\_Opening}, \textsf{stmlt\_gear\_outgoing},
    \textsf{stop\_stmlt\_general\_EV},
    \textsf{stop\_stmlt\_door\_opening},
    \textsf{stop\_} \textsf{stmlt\_gear\_outgoing}.
    \end{itemize}
\item all the events of the last family (\textit{sensing events}) are in the physical part.  
\end{itemize}

As a matter of fact, it is possible to define a systematic process to
guide the decomposition process.

%===========================================================

\section{Discussion: Coverage of the Requirements and Assessment}
% Finally, a discussion on the applicability and efficiency of the set up formal method(s) should be provided.
\label{section:discussion}
% -*- coding: utf-8 -*-
%
%-----------------------------
% Landing System ABZ'2014
% 
%\section{Discussion}

We begin the section with the analysis of the experimental results;
then we discuss the coverage of our study and
experimentation. Methodological shortcomings are commented and we
finish by generalising our approach to help the interested readers to
reuse our work in similar case studies.

\subsection{Experimentation with Rodin and statistics}

\subsubsection{Rodin}
The Rodin tool is very efficient for proving the Event-B models; a
very high percentage ($\sim$ 90\%) of proof obligations was
automatically discharged.  Note that the current version of the
Event-B models is partial as we focus on representative events instead
of being exhaustive specifications. The specifications are available
on a
website\footnote{\url{http://www.lina.sciences.univ-nantes.fr/aelos/softwares/LGS/}}.

\begin{table}
\begin{center}
\begin{tabular}{|l|l|l|l|l|l|}
\hline
 ~& Total & Auto & Manual & Reviewed & Undisch. \\
\hline
LandingSys5 & 619 & 547 & 6 & 0 & 66 \\
\hline
\multicolumn{6}{|c|}{\textbf{Abstract model}}\\
\hline
\hline
Landing\_DP\_Ctx & 0  & 0  & 0  & 0  & 0 \\
\hline
LandingSysDP\_A & 115  & 114  & 1  & 0  & 0\\
\hline
LandingSysDP\_SWITCH\_A  & 5  & 3 &  0  & 0  & 2\\
\hline
LandingSysDP\_DOOR\_A  & 42  & 42  & 0 &  0  & 0 \\
\hline
LandingSysDP\_DOOR\_GEAR\_A  & 79  & 79  & 0 &  0 &  0 \\
\hline
LandingSysDP\_DOOR\_GEAR\_TIME\_A  & 2  & 2  & 0  & 0  & 0 \\
\hline
\multicolumn{6}{|c|}{\textbf{Models of the vertical refinement}}\\
\hline
\hline
LandingSysDP\_DGT\_R1\_In  & 52 &  50 &  0  & 0  & 2 \\
\hline
LandingSysDP\_DGT\_R2\_INOUT  & 56  & 56  & 0  & 0  & 0 \\
\hline
LandingSysDP\_DGT\_R3\_INOUTDOOR  & 128  & 81 &  5 &  0  & 42 \\
\hline
LandingSysDP\_DGT\_R3INOUTDOORGEAR  & 140  & 120  & 0  & 0  & 20 \\
\hline
\end{tabular}
\end{center}
\caption{Statistics of PO generated and proved with Rodin}
\label{table:POstats}
\end{table}

Statistics on Proof Obligations are given in
Tab.~\ref{table:POstats}. From a total of 619 POs, 547 of them were
automatically discharged by Rodin and 6 of them were interactively
discharged. Most of the POs at the abstract levels were proved. The
undischarged POs are related to the structural refinement and
specifically they are related to the binding invariants. 

\medskip However the Rodin tool is inefficient with large models; it
lacks of space and becomes very slow and unpractical. Therefore
managing very large models requires a rigorous slicing and several
small steps of refinements. This is the reason why we have introduced
many refinements, but it is still not enough, the slicing should be
more fine.

As far as the ProB animation tool (integrated in Rodin) is concerned,
it is very helpful to tune the Event-B models; but unfortunately for
the large models as the ones built (several finite sets, several
variables and several events) for the landing system the animation
tool is inefficient. It lacks of space and results in frozen work
space. This is cumbersome as unfortunately we was not able to animate
the last steps of our development. In a similar way, the ProB model
checker was only able to exploit the preliminary abstract models; but
as soon as we introduce more complex state space and events, we were
not able to model-check the B models.

\subsection{Coverage and assessment}

The proposed Event-B specification presented in this article covers
the main aspects of the landing system: the digital part with modules
redundancy, its physical part (mechanical and hydraulic environment
and pilot interface) and their interactions.

We have emphasised a treatment of the global features of the landing
system. Therefore we have dealt with all the aspects of the control of
the physical part, starting from an handle action on the pilot
interface. Only the \textit{outgoing} sequence is treated here. The
second sequence should be treated in a similar way. Each step of the
interaction between the digital part are treated: we have a cycle to
sense the environment, to make a decision, to give an order. This
cycle is repeatedly observed.

In Table~\ref{table:Reqstats} we provide a synthesis of the coverage of
the requirements. The presented work covers mainly the safety
properties; liveness properties are treated by adapting Lamport's
logical clocks~\cite{DBLP:journals/cacm/Lamport78}; nevertheless we
have not deal with time constraints. In our study of the landing
system we have considered representative properties. Indeed the
requirement document lists at pages 18 and 19, two categories of
properties: normal mode and failure mode requirements. But in each
category the stated requirements are quite similar.

\begin{table}
\begin{center}
\begin{tabular}{|l|c|c|}
\hline
Category & Covered requirements & Uncovered requirement \\
\hline
\hline
 \textbf{Safety} &  $R_2$, $R_3$, $R_4$ and $R_5$  & \\
\hline
 \textbf{Liveness} & $R_1$  &  $R_{1bis}$\\
\hline
 \textbf{Nonfunctional} &  & $R_6$, $R_7$ and $R_8$ \\
\hline
\end{tabular}
\end{center}
\caption{Coverage of the requirements}
\label{table:Reqstats}
\end{table}

Code generation was out of our solution. We build on the experiments
of several case studies where the Event-B was used and where some
methodological guidelines was
provided~\cite{DBLP:conf/icfem/DamchoomBA08,DBLP:conf/sbmf/DamchoomB09}.
Accordingly, feature augmentation and structural refinement methods
appear very efficient.

\subsection{Methodological shortcoming}

\begin{figure}[htbp]
\centering
\includegraphics[width=.6\linewidth]{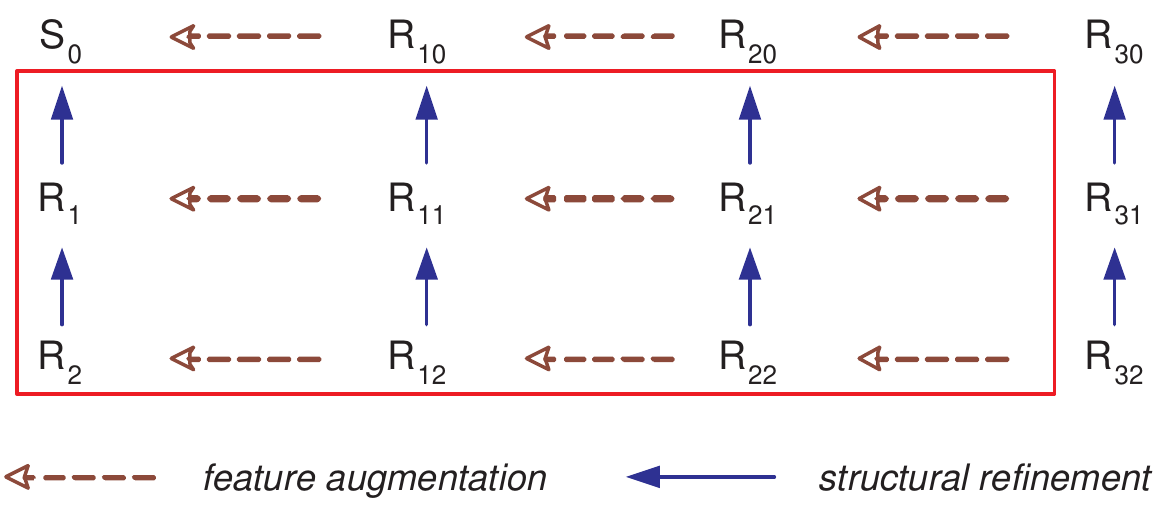}
\caption{Combining horizontal and vertical refinements}
\label{figure:refinements}
\end{figure}

One flaw of the Event-B top-down approach is the constraint imposed by
the evolution of the global abstract model defined before its
refinement to the concrete models.  For example in
Fig.~\ref{figure:refinements}, $M_{30}$ stands for the global abstract
model; only the models of the first horizontal line (indexed by
{\small 10,20,...}) and the last column (indexed by {\small 30}) are
pertinent during the specification (the models that are inside the box
are not considered.  This constraint prevents for an incremental model
evolution.  Indeed, if we miss some features in the abstract state, we
will have to reconsider completely the structural refinements.  It
would be interesting to be able to mix both horizontal and vertical
refinements in an incremental view of the design method as shown by
the intermediary steps inside the box of
Fig.~\ref{figure:refinements}. In~\cite{DBLP:conf/zum/Back02} Back,
have proposed guidelines for this purpose; an adaptation of this work
to Event-B is likely to be interesting.

The reuse of existing independent models, with a bottom-up approach,
would be interesting for managing large Event-B models. A typical
example is the composition of existing models to build a given
abstract model where each part can be modified and refined separately.

\subsection{A Generic approach for a control system}

Many features of our solution emphasise the genericity of the used
approach. During our work on the landing system, which is viewed as
representative of critical embedded control systems, we pay attention
to emphasise the features that can be reused in similar case studies.

\begin{figure}[htbp]
\centering
\includegraphics[width=.5\linewidth]{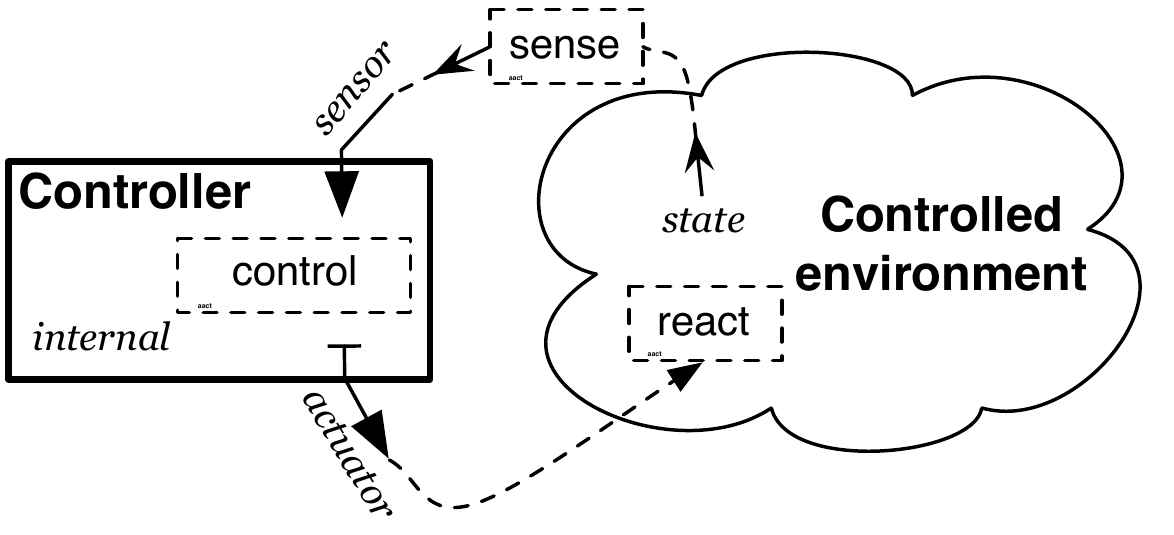}
\caption{Generic shape for Event-B based model of control system}
\label{figure:controlSysPattern}
%
% en fait, monitor = control
%
\end{figure}

Roughly, Fig.~\ref{figure:controlSysPattern} depicts a general
principle that may govern the organisation of event-based models of
control systems. We consider  that the high level state space of the
control system can be  described on the basis of the
elicitation of the interface variables between the control part and
the physical part of the considered systems.

The dashed ovals are representative of the parametric events; they are
linked to  both sensing and control family of events. 
They should be replaced by the effective events related to
the logic of a specific case study. For example in the case of the
landing system the merging event is a logical OR; in a different case
study the merging may correspond to another specific policy.

\begin{figure}[htbp]
\centering
\includegraphics[width=.4\linewidth]{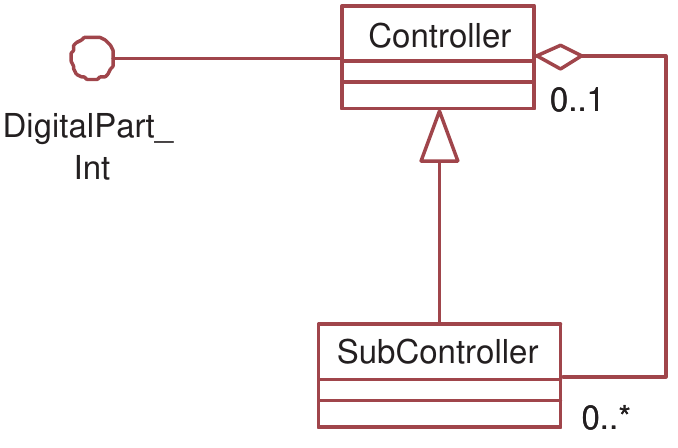}
\caption{Incremental refinement pattern of duplicated controllers}
\label{figure:digitalPart_pattern}
\end{figure}

The case where the internal control modules (sub-controllers) have the
same interface as the main control module (the controller) is a
specific case; it is structured by the meta-model in
Fig.~\ref{figure:digitalPart_pattern}.

Additional specific features and modelling guidelines are the
following.
\begin{enumerate}
\renewcommand{\theenumi}{\textit{\roman{enumi}})}%
\item \textit{Description of the interface variables of the
    controller:} the interface variables into three categories (input
  variables, output variables and internal variables); each category
  of variables are set by the events which are classified in both
  categories: sensing, or controlling.  These event families are
  conform with the \textsf{sense/decision/control} control cycle.
\item Use of \textit{feature augmentation to bind the controlled
    environment: } this is achieved on the basis of the sensing
  events, which in turn need the description of the controlled
  environment. Abrial's advices on proceeding with small steps of
  refinements are very helpful here. Instead of a big step of
  refinement including several variables and events, several small
  steps of refinements dedicated to variables and events, in an
  incremental way, are more efficient.
\item \textit{Reachability property with partial ordering:} specific
  events (not exactly at the same granularity with the B model events)
  with timestamps are systematically used to order and to reason on
  reachability properties.
\item Use of \textit{Structural refinements} based on the control
  events to refine the controller. The involved categories of
  variables are the internal variables; the input variables are
  spawned inside submodules if any; in the same way output variables
  should be updated by promotion from the submodules if any.
\begin{figure}[htbp]
\centering
\includegraphics[width=.8\linewidth]{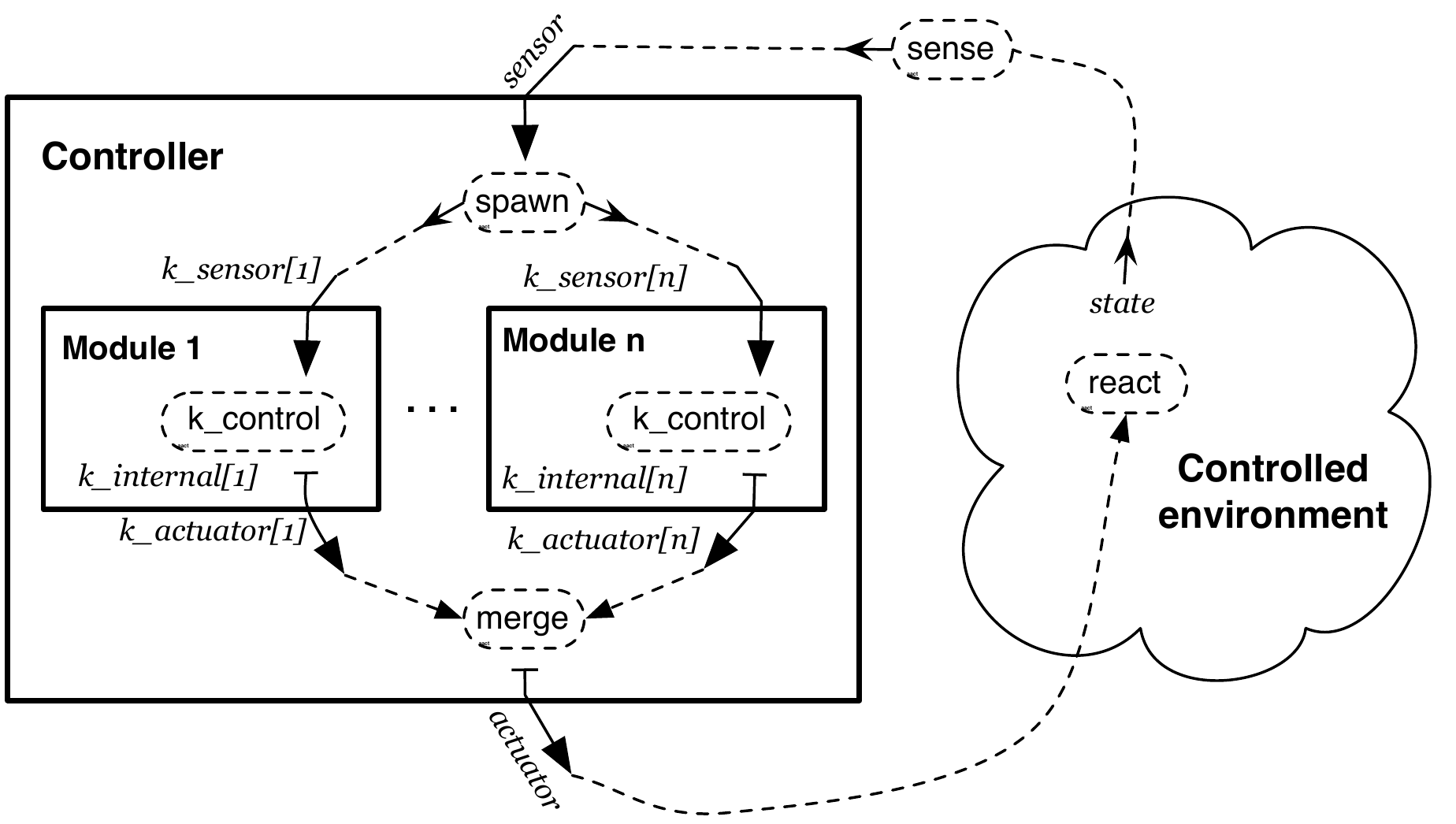}
\caption{Generic Event-B based control system with modules redundance}
\label{figure:controlSysPattern2}
\end{figure}

\begin{figure}[htbp]
\centering
\includegraphics[width=.8\linewidth]{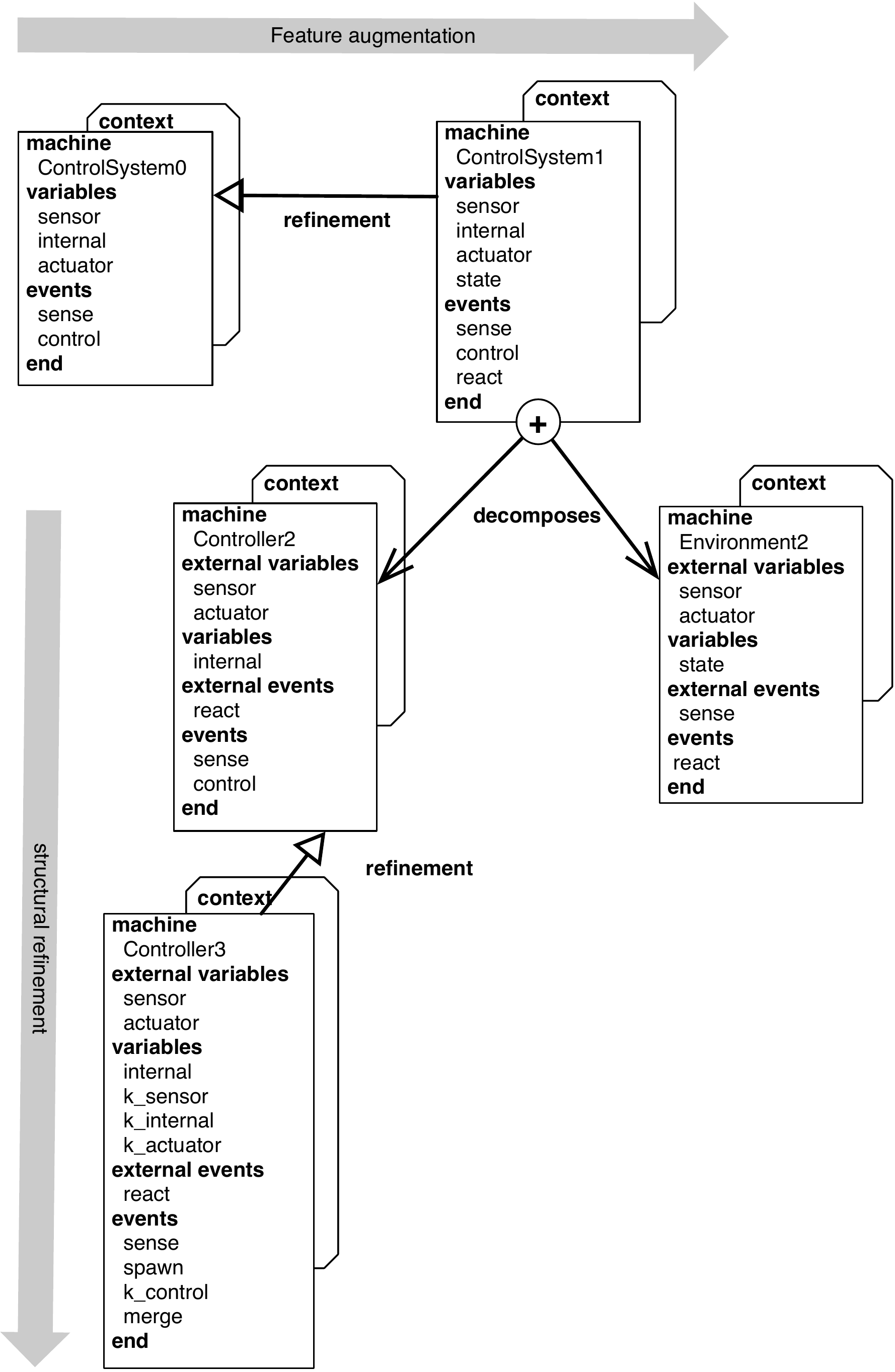}
\caption{Generic Event-B models}
\label{figure:pattern:systematicDecompo}
\end{figure}

\item \textit{Composition of several redundant blocks:} when a
  controller is made of several redundant modules, it is
  straightforward to describe a generic module and use an indexing
  function to compose several instances of such modules (see
  Fig.~\ref{figure:controlSysPattern2}).
  \begin{itemize}
  \item \textit{Immersing variables inside modules:} the values coming
    from outside one or several modules can be systematically immersed
    inside the modules with an event such as $spawn\_gear\_extended$.
  \item \textit{Promoting variables outside a module:} in a symmetric
    way, the values going outside a module or several modules are
    systematically described using a promotion pattern described in
    Sect.~\ref{section:refineDigital} for merging the output variables
    of the internal computing modules.
  \end{itemize}
  When the modules are not redundant, each one should be refined
  conveniently, but the described treatment of the inputs and outputs
  is the same.
\end{enumerate}

Figure~\ref{figure:pattern:systematicDecompo} shows Event-B patterns
at each ---horizontal and vertical--- refinement level, from the most
abstract model \textsf{ControlSystem0} describing only the controller
interface to the systematic decomposition into two parts:
\textsf{Environment2} and \textsf{Controller3}.

\section{Conclusion}
\label{section:conclu}
% -*- coding: utf-8 -*-
%
% Landing System
% ABZ'2014 
%\section{Conclusion}
%

%Bilan
We presented a complete study of the landing gear system from the
point of view of the modelling and the verification of given
properties. It is a contribution to the specification challenge
submitted to the ABZ communities. We used Event-B and the related
horizontal and vertical refinement approaches. An important part of
the requirement document has been treated by considering both the
digital part and the physical part. The digital part has been refined
until its decomposition into the two redundant modules introduced in
the requirement. Code generation is not attacked.

 After the requirement capture where we used several approaches, we
proceed by a systematic approach which can be reused in similar
studies of reactive embedded system. Describing precisely the
interface between the digital part and its environment (pilot
interface, physical devices) is an important starting point to build
the abstract model of the control system. The variables defining the
interface are systematically classified into four categories which
structure the model of the control system and its further
refinements. We have defined some event description patterns related
to the four variable categories which structure the control
system. Families of events are identified to structure the modelling
of the interaction between the digital part and its environment. The
events needed to build and refine the digital part depends on the
interface variables and they are systematically described by
considering the standard \textsf{sense/decision/order} control
cycle. It appears that each family of events is precisely located
either in the horizontal refinement or in the vertical
refinement. Consequently our approach is systematic; it can be reused
with the provided guidelines and it can be assisted by tools. Moreover,
we have proposed a new approach to deal with reachability properties;
its is based on Lamport's logical
clocks~\cite{DBLP:journals/cacm/Lamport78} and the partial ordering of
specific events introduced in addition to the Event-B events.

%Lessons learnt 
Many lessons was drawn from our experimentation. The use of Event-B is
relevant to attack the landing system. Indeed few properties related
to time constraints and reachability are not deal with. We have not
spent time to extend the method to overcome this flaw. But we have
shown that, with some crafty modelling approaches we can come up with
a part of these reachability properties. This opens the way for
improving the Event-B method and enriching its tools.

%Perspectives
We plan to generalise the generic approach presented here and build a
related tool to help in modelling and structuring the development of
embedded control systems.  The proposed approach muse be compared to
the four relational variables model proposed by
Parnas~\cite{DBLP:journals/scp/ParnasM95} for rigorous system
development.  The approach we have used to deal with reachability
properties will be studied in more details and extended to tackle more
properties.

\bibliographystyle{splncs}
\bibliography{biblio}

\end{document}